\documentclass[prb,  twocolumn, showpacs, superscriptaddress,longbibliography]{revtex4-1}
\usepackage{amsmath, amsfonts, amssymb, mathrsfs}
\usepackage{graphicx, color}

\usepackage[caption=false]{subfig}
\usepackage{bm}
\usepackage{braket}
\usepackage{epstopdf}
\usepackage{dsfont}
\usepackage{float}
\usepackage[dvipsnames]{xcolor}
\usepackage{listings}
\usepackage{program}
\usepackage{mathtools}
\usepackage{hyperref}
\hypersetup{
   colorlinks,
   citecolor=blue,
   filecolor=blue,
   linkcolor=blue,
   urlcolor=blue,
   breaklinks=true
}

\definecolor{grey}{rgb}{.6,.6,.6}

\newcommand{\ti}[1]{{\emph{#1.}}}

\newcommand{\norm}[1]{\left\|#1\right\|}

\newcommand{\overbrack}[1]{\overbracket[0.5pt][1.5pt]{#1}\!} %line width 0.5pt, bracket height 1.5pt
\newcommand{\tid}{\text{Id}}

\begin{document}
\title{$\mathcal{L}^2$ localization landscape for highly-excited states}
\author{Lo\"ic Herviou}
\affiliation{Department of Physics, KTH Royal Institute of Technology, Stockholm, 106 91 Sweden}
\author{Jens H. Bardarson}
\affiliation{Department of Physics, KTH Royal Institute of Technology, Stockholm, 106 91 Sweden}
\begin{abstract}
The localization landscape gives direct access to the localization of bottom-of-band eigenstates in non-interacting disordered systems.
We generalize this approach to eigenstates at arbitrary energies in systems with or without internal degrees of freedom by introducing a modified $\mathcal{L}^2$-landscape, and we demonstrate its accuracy in a variety of archetypal models of Anderson localization in one and two dimensions.
This $\mathcal{L}^2$-landscape function can be efficiently computed using hierarchical methods that allow evaluating the diagonal of a well-chosen Green function.
We compare our approach to other landscape methods, bringing new insights on their strengths and limitations.
Our approach is general and can in principle be applied to both studies of topological Anderson transitions and many-body localization.

\end{abstract}
\maketitle
\ti{Introduction}---The theoretical discussion of Anderson localization, the strict confinement of matter waves to a finite subspace due to destructive quantum interference, dates back to 1958\citep{Anderson1958}.
Progress since then has come in bursts, often separated by long intervals.
Only after over twenty years did mathematical proofs start to appear~\citep{Goldshtein1977, Kunz1980} and the scaling theory of localization was introduced\citep{Abrahams1979}, suggesting that all eigenstates in low dimensions are localized.
Two major recent modern developments involve the interplay of localization with topology and interactions:
Surfaces of topological insulators\citep{Hasan2010} resist localization\citep{Schnyder2008, Evers2008, Ludwig2015} and extended bulk states are obtained at the transition between two topologically distinct insulating phases\citep{Brouwer1998, Senthil1998, Gruzberg1999, Senthil1999, Read2000, Titov2001}.
Interactions give rise to many-body localization, in which an eigenstate phase transition is obtained at energies high above the ground state\citep{Altshuler1997, Gornyi2005, Basko2006, Abanin2017, Alet2018}.
These phenomena only started to be understood in the last couple of decades.

One reason for this slow progress may be that localization is due to nontrivial interference patterns that are not easily guessed from the random potential the particles move in.
There is generally no obvious correlation between the localization centers of wave functions and the potential extrema.
In a sense, this means that there is no obvious classical starting point from which one can do simple perturbation theory.
Now, in a series of fascinating work, such a starting point may have been identified in the so-called localization landscape\citep{Filoche2012, Filoche2013, Lyra2014, Arnold2016, Steinerberger2017, Filoche2017, Piccardo2017, Li2017, Arnold2019}. 
The localization landscape is an effective potential obtained from the initial random potential, and it has the property that its peaks and valleys predict the location of the few lowest energy localized wave functions.
It furthermore gives the correct integrated density of states\citep{David2019} at low energy from a simple Weyl law, which otherwise badly fails when using the original potential.

The original formulation\citep{Filoche2012} of the localization landscape is for scalar field theories with a real and positive Green function, and applies strictly only to low energy states close to the bottom of the energy spectrum.
These constraints prevent direct applications to many of the modern approaches mentioned above, where the interesting physics often takes place in states at or near the middle of the spectrum.
Here, we introduce an extension of the localization landscape, which we coin the $\mathcal{L}^2$-landscape, that faithfully captures the localization of eigenstates at all energies and in the presence of internal degrees of freedom.
The $\mathcal{L}^2$-landscape can be \emph{efficiently} numerically obtained in generic physical models, in the absence of long-range hopping.
We exemplify its validity and reliability through several archetypal models of localization in one and two dimensions.
This new landscape is applicable to both topological models and many-body Hamiltonians and can therefore be used to analyze most localization problems.

An alternative extension of the localization landscape to Dirac fermions was recently introduced by Lemut \textit{et al.}\ in Ref.~\onlinecite{Lemut2019}.
This method, based on the comparison-matrix\citep{Ostrowski1937, Ostrowski1956}, has the advantage that is retains the simplicity of the original landscape, and can be applied to Dirac Hamiltonians with inner degrees of freedom.
Neither the original landscape nor the one based on the comparison matrix can, however, describe a generic high-energy state as does our $\mathcal{L}^2$-landscape, albeit at the cost of a slightly reduced efficiency. 
We conclude our work by briefly comparing our method to these alternatives, bringing insights into the strengths and weaknesses of conventional localization landscape approaches.

\ti{$\mathcal{L}^2$ localization landscape}---In their original paper\citep{Filoche2012}, Filoche and Mayboroda considered the localization of a scalar field, or equivalently of spinless fermions.
Let $H$ be the corresponding single-particle Hamiltonian and $\Ket{\phi^\beta}$ an eigenstate of $H$ with eigenvalue $E^\beta$. 
We denote by $\phi^\beta_j=\Braket{j \vert \phi^\beta}$ its amplitude at site $j$.
By application of the inverse of the Hamiltonian, one straightforwardly obtains
\begin{align}
\lvert \phi^\beta_j \rvert &= \lvert E^\beta \sum\limits_m (H^{-1})_{j, m} \phi^\beta_m \rvert \label{eq:LocLand-equality} \\
&\leq \lvert E^\beta \rvert\ \norm{\phi^\beta}_\infty \sum\limits_m \lvert (H^{-1})_{j, m} \rvert  \equiv \lvert E^\beta \rvert\ \norm{\phi^\beta}_\infty  u_j.\label{eq:LocLand-inequality}
\end{align}
$u$ is called the localization landscape.
The key insight of Ref.~\onlinecite{Filoche2012} was to realize that in a wide class of models, $H^{-1}$ can have all components positive, implying that $u$ is a solution to the differential equation
\begin{equation}
H u = 1.
\end{equation}
The requirement of element-wise positivity of $H^{-1}$ enforces strong restriction on $H$: it must be a monotone matrix\citep{CollatzBook}, a class of matrices that is generally hard to characterize.
In the case of a real symmetric matrix with all off-diagonal (hopping) terms negative, such as in the standard Anderson model, a necessary and sufficient condition is that $H$ is positive definite.
The localization landscape proves to tightly bound bottom-of-band eigenstates, almost saturating Eq.~\eqref{eq:LocLand-inequality}, in a wide variety of models~\citep{Filoche2012}. 
This saturation implies that the lowest-energy eigenstates are localized at the peaks of the landscape and different eigenstates are separated by landscape minima.
Indeed, we can rewrite the localization landscape as
\begin{equation}
u_j = \sum\limits_\beta \frac{ \phi_j^\beta }{ E^\beta } \sum\limits_m  \phi_m^\beta.  \label{eq:LLStruct}
\end{equation}
By construction, due to the inverse energy factor, eigenstates with the lowest energy will contribute more to the localization landscape than ones at higher energies.
On the other hand, high-energy states are not accurately localized by the landscape. 
This landscape can therefore not be used to study center-of-band properties.

We propose to overcome this limitation by slightly modifying the definition of the localization landscape.
Starting from Eq.~\eqref{eq:LocLand-equality}, we apply the Cauchy-Schwartz inequality to obtain
\begin{align}
\lvert\phi^\beta_j \rvert &\leq \lvert E^\beta \rvert \norm{\phi^\beta}_2 \sqrt{\sum\limits_n (H^{-1})_{j, n} (H^{-1})_{j, n}^* }\\
&=   \lvert E^\beta \rvert  \sqrt{(\mathcal{M}^{-1})_{j, j}},  \label{eq:LocLand-L2}
\end{align}
where $\mathcal{M}=H^\dagger H$ is a Hermitian positive definite matrix and we assume normalized eigenfunctions with $\vert \vert \phi^\beta\vert \vert_2=1$.
The $\mathcal{L}^2$-landscape $u^{(2)}$ is then defined by
\begin{equation}
u^{(2)}_j= \sqrt{(\mathcal{M}^{-1})_{j, j}}. \label{eq:LocLand-L2-def}
\end{equation}
$\mathcal{M}$ is invertible as long as $H$ is invertible, and the inequalities are valid whether $H$ is Hermitian or non-Hermitian.
The largest contributions to the landscape $u^{(2)}$ are from the eigenstates with the smallest absolute energy.
With this definition, there is no requirement that $H$ be positive definite, and we can therefore explore localization at all energies by simply shifting the Hamiltonian by a constant real factor $E_0$.
Note also that the normalization by the largest element of $\phi^\beta$ has vanished, replaced by its 2-norm (equal to $1$ by convention).
The change in normalization can conveniently help to differentiate localized and delocalized regimes.
In the original formulation, several tightly localized but close-in-energy eigenstates would have exactly the same landscape signatures as a state delocalized on a subpart of the system (with well separated peaks) as the difference in amplitude of the wave functions is not taken into account.
Eq.~\eqref{eq:LocLand-L2-def} is valid in the continuum limit, and can be straightforwardly applied to systems with internal degrees of freedom.

To ensure that $\mathcal{M}$ can be inverted, it is convenient to introduce a complex energy shift $\varepsilon$ and work with the matrix $\tilde{H}=H+i\varepsilon \, \mathrm{Id}$.
The energy in the bound is then renormalized to $E_\varepsilon^\beta=\vert E^\beta+i\varepsilon\vert=\sqrt{(E^\beta)^2+ \varepsilon^2}$.
$\varepsilon$ can be taken as small as required (though too small a value may affect the coordination number of $\mathcal{M}$ and therefore the numerical precision of certain computations), and needs to be smaller than the level spacing at the probed energy range in order to resolve different eigenstates.
We can gain an intuition for this by writing, for an Hermitian Hamiltonian, the square of the landscape as
\begin{equation}
(u^{(2)}_j)^2 = \sum\limits_\beta \frac{\vert \phi^\beta_j \vert^2}{(E^\beta-E_0)^2 + \varepsilon^2},
\end{equation}
where we have now also explicitly included the real energy shift $E_0$.
%This form of the landscape makes explicit the similarity with the local density of states $\rho_j(E) = \sum\limits_\beta \vert \phi^\beta_j \vert^2 \delta (E - E^\beta) $ as
We therefore have that 
\begin{equation}
	\label{eq:limitLDOS}
\varepsilon (u^{(2)}_j)^2 \xrightarrow[\varepsilon \rightarrow 0]{} \rho_j(E_0),
\end{equation}
where $\rho_j(E) = \sum\limits_\beta \vert \phi^\beta_j \vert^2 \delta (E - E^\beta)$ is the local density of states at site $j$ and energy $E$.
This explains why the $\mathcal{L}^2$-landscape provides an efficient description of states close to $E_0$, while the presence of the factor of $\varepsilon$ on the left hand side of relation~\eqref{eq:limitLDOS} means that states further away from $E_0$ also contribute to the landscape. 
%This explains why the $\mathcal{L}^2$ landscape provides an efficient description of states within a distance $\varepsilon$ of $E_0$, while the presence of the factor of $\varepsilon$ on the left hand side of relation~\eqref{eq:limitLDOS} means that states further away from $E_0$ also contribute to the landscape. 

The $\mathcal{L}^2$ localization landscape can be computed efficiently, even if it does not satisfy a simple (discrete) differential equation.
Indeed, for short ranged Hamiltonians, numerous methods have been developed to compute the diagonal of the Green functions efficiently, such as hierarchical algorithms\citep{Buzbee1970, George1973, Sancho1985, Svizhenko2002, Lewenkopf2013} (that can also take advantage of the positive definiteness of $\mathcal{M}$).
More refined algorithms in two dimensions\citep{Li2008, Lin2009, Li2011} can compute the diagonal of the inverse in $\mathcal{O}(L^3)$ operations, where $L$ is the linear dimension of the two-dimensional system.
Moreover, several methods\citep{Robinson1992, Golub1994, Benzi1999} exist to numerically derive upper bounds on the components of the inverse of Hermitian definite positive matrices, that can readily be applied here.\\

\begin{figure}[tb!]
\includegraphics[width=\linewidth]{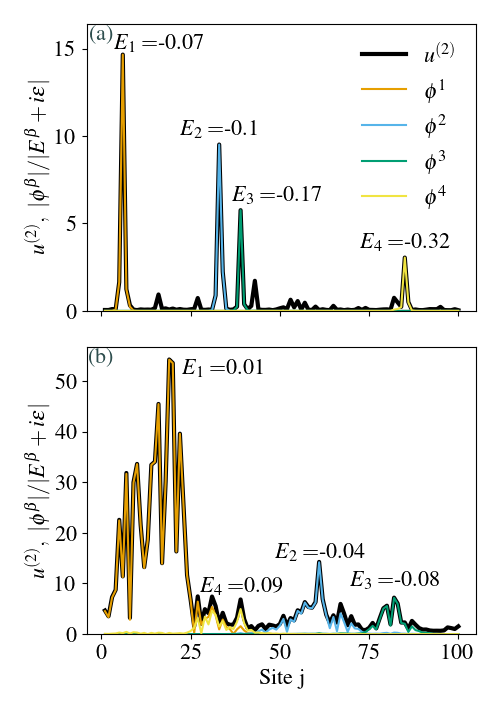}
\caption{
$\mathcal{L}^2$-landscape and the four eigenstates closest to zero energy ($E_0=0$) in the Anderson model for disorder strengths $W=25$ (a) and $W=2$ (b). 
The eigenstates are normalized by their energy and $\varepsilon = 10^{-3}$ is fixed to be smaller than the typical mean level spacing.
The different peaks in the localization landscape coincide with the different eigenstates and their location. 
The low minima form domain bounds that separate different eigenstates at low-energy. 
$u^{(2)}$  predicts accurately the localization and ordering of the states in all cases, and tightly bounds the localization of these states. 
}
\label{fig:AndersonComparison-L2}
\end{figure}

\ti{Anderson model}---We first illustrate our method in the prototypical one-dimensional Anderson model for localization, with Hamiltonian 
\begin{equation}
H = -t\sum\limits_j (c^\dagger_j c_{j+1} + c^\dagger_{j+1} c_j) + \sum\limits_j V_j c^\dagger_j c_j. \label{eq:AndersonModel}
\end{equation}
$c_j$ ($c_j^\dagger$) is the fermionic annihilation (creation) operator on site $j$, $t$ is the hopping amplitude (set to $1$ in the following) and $V_j$ is a random on-site potential uniformly distributed in $[-W, W]$.
An arbitrarily weak disorder is enough to localize all eigenstates at all energies in the thermodynamic limit, including in the middle of the spectrum.
In Fig.~\ref{fig:AndersonComparison-L2} we show the $\mathcal{L}^2$ localization landscape at zero energy in a chain of $L = 100$ sites, and compare it with the few eigenstates nearest in energy.
Taking the cut-off $\varepsilon$ to be smaller than the typical level spacing, $u^{(2)}$ accurately describes the localization of the states close to $E_0=0$ at both strong and weak disorder.
As with the conventional landscape, many eigenstates are captured by a single computation of the landscape, whether at strong or weak disorder.
The ordering of peak amplitudes matches the eigenstate ordering.\\

\begin{figure}[tb!]
\begin{center}
\includegraphics[width=\linewidth]{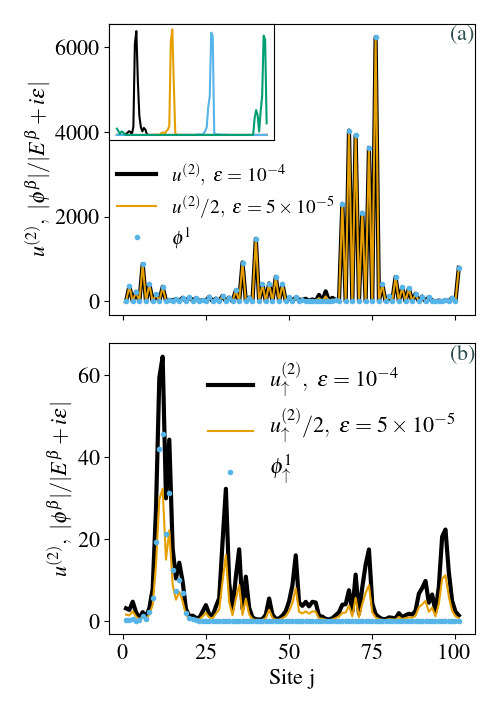}
\end{center}
\caption{
The $\mathcal{L}^2$ localization landscape at $E_0=0$ for two values of $\varepsilon$ and the lowest lying eigenstate in the one-channel (a) and two-channel (b) chiral Anderson model, for $V_0=4$ and $L=101$ sites. 
We only plot the spin up component of the landscapes and wave functions in the two-channel case for simplicity; the other component can be obtained by symmetry. 
For one channel, there exists an extended zero mode which gives a clear contribution to the landscape, with an amplitude that scales as $\varepsilon^{-1}$. 
Conversely, the part of the landscape that does not scale with $\varepsilon$ (e.g., around $j=60$) corresponds to higher energy states.
In the inset we show, for reference, the wave functions of the four states in the bulk of the band with $E_0=V_0$.
For two channels, there is no zero mode, and the lowest energy states are localized. 
The rescaled landscape does not match its initial counterpart. 
The landscape is a less tight bound than usual due to the chiral symmetry which doubles the number of states.
To get a tighter bound one can split the pairs of states at $\pm E$ by a weak breaking of the symmetry.
}
\label{fig:CA}
\end{figure}

\ti{Chiral Anderson model}---The ability to access arbitrary energies allows us to access more refined properties of localization, such as due to the presence of symmetries.
In one dimension, the presence of chiral symmetry leads to an even-odd effect in terms of the number of channels\citep{Evers2008, Morimoto2015}; indeed, due to the symmetry, states either come in pairs $(E, -E)$ or have zero energy.
For an odd number of channels (and an odd number of sites at finite sizes) there therefore must exist a symmetry protected zero energy eigenstate.
This zero energy state is delocalized even in the presence of strong disorder.
In Fig.~\ref{fig:CA}\href{fig:CA}{a}, we compare the $\mathcal{L}^2$ localization landscape and eigenstates of the minimal single-channel chiral Hamiltonian
\begin{equation}
H=- \sum\limits_j t_j c^\dagger_j c_{j+1} + h.c.
\end{equation}
with $t_j$ taken uniformly in $[-V_0, V_0]$.

Though the landscape may appear similar to the one obtained in Fig.~\ref{fig:AndersonComparison-L2}, we can identify that most peaks are contributions of a zero mode by varying the cutoff $\varepsilon$.
Indeed, the energy in Eq.~\eqref{eq:LocLand-L2} is given by $\vert E^\beta + i \varepsilon \vert$.
When $\varepsilon \rightarrow 0$, the contributions to the landscape of states with nonzero energy are suppressed compared to the divergent contribution of the zero energy eigenstates and we can identify the zero mode contributions by computing the landscape for two different cut-offs: bounds of the zero modes will scale as the inverse of $\varepsilon$.
In Fig.~\ref{fig:CA}, we show an example with such a delocalized state.
Additionally, around $j=60$, one can see a few peaks where the landscape does not scale linearly with $\varepsilon$; this is where the first excited states are localized.
In the absence of degeneracies, it is then immediate to identify that the zero mode spans large part of the system.
One can verify by shifting the energy reference $E_0$ that bulk states are localized.

Conversely, in the case of an even number of channels, the symmetry no longer guarantees the presence of a zero mode, and the eigenstates close to zero energy are all localized.
The Hamiltonian
\begin{equation}
H=- t\sum\limits_j \vec{c}_{j}\:\!^\dagger\sigma^z \vec{c}_{j+1} + h.c. - \sum\limits_j V_j \vec{c}_{j}\:\!^\dagger\sigma^y \vec{c}_{j+1} + h.c.
\end{equation}
with $\vec{c}=(c_\uparrow, c_\downarrow)$ two fermionic species, $\sigma^\alpha$ with $\alpha=x,y,z$ the Pauli matrices  and $V_j \in [-V_0, V_0]$, is an example of two-channel chiral Anderson model.
The chiral symmetry is realized by
\begin{equation}
\sigma^x H \sigma^x = - H
\end{equation}
As shown in Fig.~\ref{fig:CA}\href{fig:CA}{b}, there are no zero modes and the eigenstate closest to zero energy is now localized.
The localization landscape bounds the eigenstates less tightly than in the previous examples due to the chiral symmetry: states comes in pairs of opposite energies which have exactly the same renormalized energies and similar local polarization.
These two contributions therefore sum up constructively and strongly relax the usual tightness of the bound.
This can be remediated by a small breaking of the chiral symmetry with a nonzero $E_0$ smaller than the mean level spacing.\\

\begin{figure}[tb!]
\includegraphics[width=0.95\linewidth]{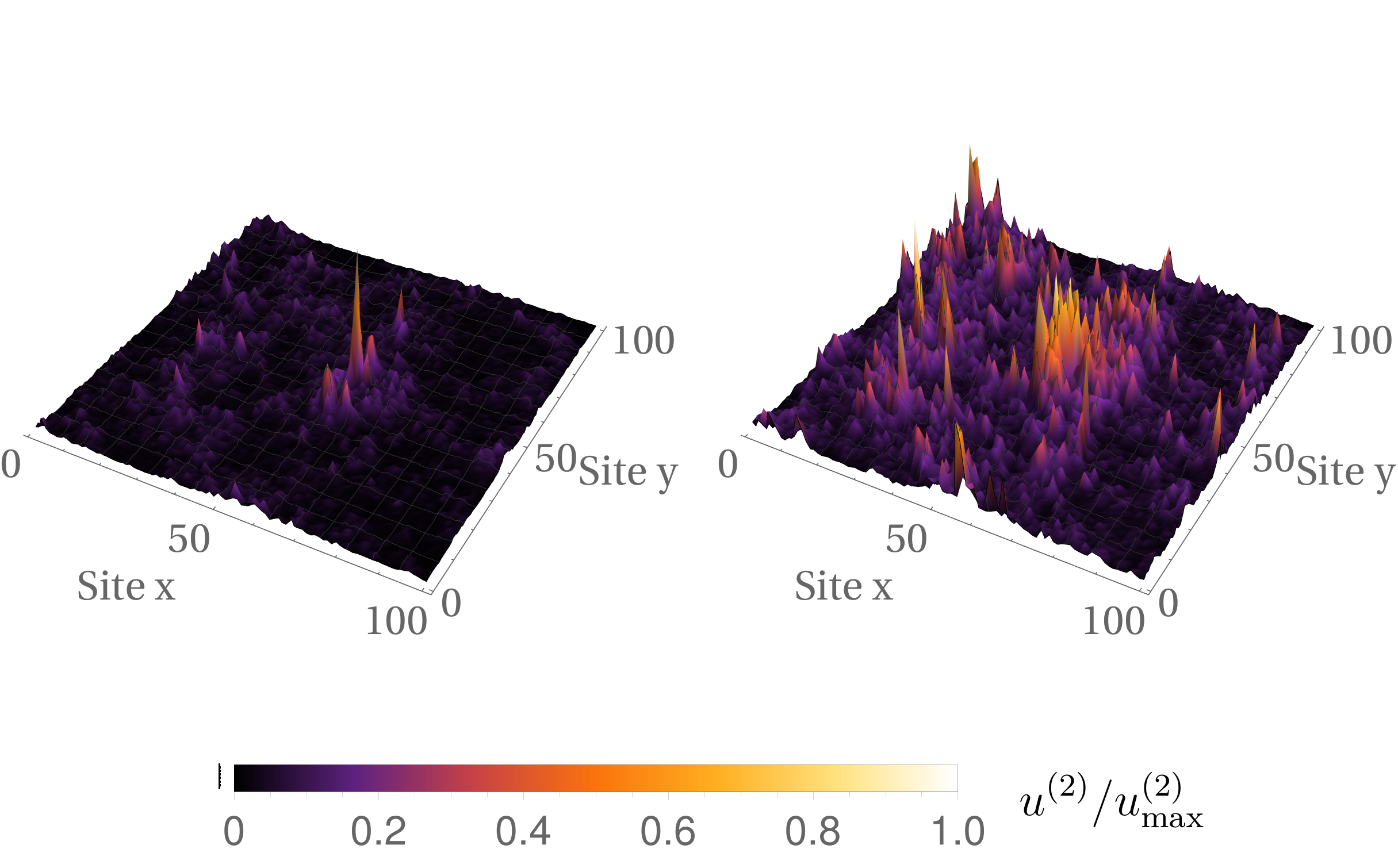}
\caption{$\mathcal{L}^2$ localization landscape for the critical two-dimensional Chern insulator in the presence of all types of disorder for both spin components (left: spin up, right: spin down). 
We fix $V_0^\alpha$ to $2t$.  
In both graphs, the vertical component depicts the low-level eigenstates, while the colorscale is the corresponding normalized value of the landscape. 
Peaks and valleys in the landscape match the lowest energy eigenstates.}
\label{fig:2DAnderson}
\end{figure}

\ti{Dirac fermions in two dimensions}---Finally, we demonstrate that the $\mathcal{L}^2$-landscape also captures the (absence of) localization of Dirac fermions in two dimensions.
Single Dirac cones with time reversal are not localized at any energy\citep{Bardarson2007, Nomura2007, Ostrovsky2007, Evers2008}, and belong to different universality classes depending on the form of the disorder.
A convenient lattice model to simulate a single Dirac cone is a critical two-dimensional Chern insulator on a square lattice
\begin{align}
H=&-t\sum\limits_{\langle \vec{r},  \vec{r}' \rangle} \left(\vec{c}_{\vec{r}}\:\!^\dagger \sigma^z  \vec{c}_{\vec{r}'} + h.c. \right) - \mu\sum\limits_{\vec{r}} \vec{c}_{\vec{r}}\:\!^\dagger \sigma^z  \vec{c}_{\vec{r}}\\
 &+ \Delta_x \sum\limits_{\vec{r}} \left(i \vec{c}_{\vec{r}}\:\!^\dagger \sigma^x  \vec{c}_{\vec{r}+\vec{e}_x} + h.c. \right)\\
 &+ \Delta_y \sum\limits_{\vec{r}} \left(i \vec{c}_{\vec{r}}\:\!^\dagger \sigma^y  \vec{c}_{\vec{r}+\vec{e}_y} + h.c. \right)
\end{align}
$t$, $\Delta_x$ and $\Delta_y$ act as different flavors of spin-orbit coupling, and $\mu$ is a chemical potential.
The system falls into class $D$ with the particle-hole symmetry
\begin{equation}
\sigma^x H^* \sigma^x = - H.
\end{equation}
For $\mu=\pm 4t$ and $\Delta_x$ and $\Delta_y$ nonzero, the Hamiltonian is at a critical point between a topological phase with Chern number $\pm 1$ and a trivial phase.
It presents a single Dirac cone at momentum $\vec{k}=(0, 0)$ for $\mu=-4t$ and $\vec{k}=(\pi, \pi)$ for $\mu=4t$.
We place ourselves at this phase transition and introduce all possible random local perturbations
\begin{equation}
V^a = \sum\limits_{\vec{r}, a} V_{\vec{r}}^a \vec{c}_{\vec{r}}\:\!^\dagger \sigma^a \vec{c}_{\vec{r}},
\end{equation}
with $a \in \{0, x, y, z\}$ and $V_{\vec{r}}^{a}$ taken uniformly in $[-V_0^a, V_0^a]$.
$V^0$ is a random scalar potential, $V^z$ a random mass and $V^{x/y}$ random chiral hoppings.
The random mass $V^z$ preserves the particle-hole symmetry, while the other potential terms break the symmetry such that the system falls directly into class $A$.
In class $D$, at weak disorder, the system would fall into the thermal quantum hall transition fixed point, before transitioning at higher disorder to a metallic phase, as long as the disorder averages to zero.\citep{Senthil2000, Evers2008, Bardarson2010, Medvedyeva2010, Wimmer2010}
In class $A$ on the other hand, the model flows towards the integer quantum hall transition fixed point, though with strong finite-size effects that will lead to apparent localization at strong disorder and higher-energies.\citep{Ludwig1994, Altland2002, Ostrovsky2007b, Ostrovsky2007, Evers2008}
In Fig.~\ref{fig:2DAnderson}, we compare the prediction of the localization landscape for the critical Chern insulator and the actual low-energy eigenstates in the presence of all types of disorder, for the two spin components.
Similar results are obtained in the $D$ class.
Peaks and valleys in the landscape match the ones in the eigenstates, both exactly at zero energy where the gap closes, but also deep in the band. 
We do observe the absence of localization close to zero energy, as is evident by looking at the spin-down component.\\

\ti{Discussion}---We have introduced the $\mathcal{L}^2$-landscape, an extension of the localization landscape that can be used to characterize eigenstates of a Hamiltonian in the bulk spectrum of arbitrary models.
This requires the computation of the diagonal of the inverse of the positive-definite matrix $\mathcal{M} = H^\dagger H$.
It provides an accurate and tight bound, in the absence of degeneracies, on the localization or delocalization of eigenstates at an arbitrary energy.
We have demonstrated the power of this new landscape in a variety of models in one and two dimensions, with and without internal degrees of freedom, and presenting mobility edges and other nontrivial localization properties.
In all these examples, our method successfully and accurately pinpointed the eigenstates closest to any target energy.

It is pertinent to compare our results to other landscape-based approaches.
In particular, the comparison-matrix landscape introduced in Ref.~\onlinecite{Lemut2019} can in principle be used to study states in the middle of the spectrum.
In practice, the comparison matrix needs to be positive-definite, which, in the models we considered, requires the introduction of the same shift $\varepsilon$ we introduced.
Instead of being a small control parameter, however, this shift is much larger than the mean level spacing and sometimes even of the order of the bandwidth.
The energy denominator in Eq.~\eqref{eq:LLStruct} (replacing the Hamiltonian by the comparison matrix) is then strongly flattened, with all eigenstates contributing with similar amplitudes.
The obtained landscape is then no longer a good predictor of the localization of the eigenstates closest to the target energies (see the Appendix for more details).
These are generic limitations in conventional landscape methods, as long as the energy gap to the lowest eigenstate is much larger than the level spacing.
This problem can be alleviated by certain types of disorder that make the Hamiltonian diagonal dominant, and therefore allow for small $\varepsilon$, such as discrete disorder distributions or disorder of the form $V \vec{n}.\vec{\sigma}$, with $V$ a large constant amplitude and $\vec{n}$ a random unit vector, representing strongly disordered magnetic impurities.

The generality of our approach---including both interacting systems (in configuration space for example), non-Hermitian models and continuous models---is straightforward as it only requires the invertibility of the Hamiltonian, that can be shifted by a small $\varepsilon \in \mathbb{C}$.
In particular, the possibility of targeting accurately highly excited states may prove useful for applications to many-body localization\citep{Balasubramanian2019}, though the high coordination number of the equivalent Anderson lattice may limit a purely numerical computation. 
For possible future directions, we note that wave functions at the Anderson transition point are known to exhibit multifractal behavior\citep{Ludwig1994, Janssen1994, Huckestein1995, Mudry1996, Chamon1996, Evers2000, Evers2001, Nakayama2003}. 
The properties of the critical point can be identified by computing the fractal dimension of the wave function.
How to generalize these ideas to the localization landscape, as the latter does not describe a single eigenstate, but a superposition of several with weight depending on their energies, is an interesting open question.

\acknowledgements
We thank Carlo Beenakker, Vardan Kaladzhyan, Edwin Langmann and Bj{\"o}rn Sbierski for useful discussions.
This work was supported by the ERC Starting Grant No.~679722, the Roland Gustafsson's Foundation for Theoretical Physics and the Karl Engvers foundation.
\appendix

\section{Comparison-matrix landscape method for highly excited states}

In this appendix we discuss the limitation of traditional localization landscape methods to study excited states.
For concreteness we focus on the method of Ref.~\onlinecite{Lemut2019}, which introduced a variation on the localization landscape based on the comparison matrix in order to study systems with inner degrees of freedom.
While it can in principle also be applied to middle of spectrum states, it generically fails at characterizing the localization of these states.
Here we discuss the reasons for this failure as it reveals some limitations of conventional localization landscape methods.
There exist two natural ways to study highly excited states using the comparison matrix, by introducing the Hamiltonians $H_1(\varepsilon)$ and $H_2(\varepsilon)$ defined by
\begin{equation}
H_1(\varepsilon) = H + i\varepsilon \tid,
\end{equation}
\begin{equation}
H_2(\varepsilon) = H^\dagger H + \varepsilon^2 \tid.
\end{equation}
These two Hamiltonians admit the same eigenstates as $H$ (for $H$ Hermitian) and are both invertible for $\varepsilon\neq 0$.
They satisfy Eq.~\eqref{eq:LocLand-equality} with renormalized energies given by
\begin{equation}
E_1^\beta=\sqrt{(E^\beta)^2 + \varepsilon^2} \text{ and } E_2^\beta=(E^\beta)^2 + \varepsilon^2.
\end{equation}
Both Green functions $H_{\alpha}^{-1}$ are generally not real positive, despite $H_2$ being definite positive.
In particular, $H_{1}^{-1}$ is generally complex-valued.
Ostrowski's comparison matrix\citep{Ostrowski1937, Ostrowski1956} can be introduced to solve this issue and avoid the need to compute the full inverse\citep{Lemut2019}.
The comparison matrix $\overbrack{H}$ of an Hamiltonian $H$ is defined by
\begin{equation}
 \overbrack{H}_{m, n} = 2 \lvert H_{m, m}\rvert \delta_{m, n} -  \lvert H_{m, n}\rvert.
 \end{equation}
If it is positive definite, then it verifies
\begin{equation}
\lvert H_{m, n}^{-1} \rvert \leq  \overbrack{H}^{-1}_{m, n}.
\end{equation}
We then have $\lvert\phi^\beta_m \rvert \leq E^\beta_\alpha \max\limits_n \lvert\phi_n^\beta \rvert u^\mathrm{CM}_{\alpha, m}$, where the localization landscape can be efficiently obtained by solving the equation $\overbrack{H}\,_\alpha u_\alpha^\mathrm{CM} = 1$.

The key limitation of the method, like the original landscape method, is the need for $\overbrack{H}_\alpha$ to be positive definite, and the consequences of such a requirement on $u_\alpha$.
A naive but informative sufficient condition for definite positiveness for a real symmetric matrix $A$ is
\begin{equation}
\sum\limits_m A_{m, n} > 0 \text{ for all } n.
\end{equation}
For the comparison matrix, this translates into having $\overbrack{H}$ be diagonally dominant.
This condition can always be satisfied by choosing $\varepsilon$ large enough.
On the other hand, if $\varepsilon$ is too large, i.e., much larger than the typical mean level spacing or of the order of the bandwidth, the renormalized energies $E_\alpha$ become comparable for all low-energy states.
The localization landscapes $u_\alpha$ are then no longer a good predictor of the localization of low-energy eigenstates as too many eigenstates contribute with similar amplitudes.
An alternative interpretation is that the eigenvectors of the comparison matrix are no longer close to those of the original Hamiltonian, and the landscape obtained from $\overbrack{H}_\alpha$, which describes the localization of its eigenvectors, no longer describes the eigenstates of $H_\alpha$.
Conversely, when $\overbrack{H}$ is already  diagonally dominant before introducing $\varepsilon$---for example, for well-chosen disorder distributions in the strong disorder limit---it proves to be a very efficient way to study the localization of the low-energy eigenstates.

\begin{figure}[tb!]
\includegraphics[width=\linewidth]{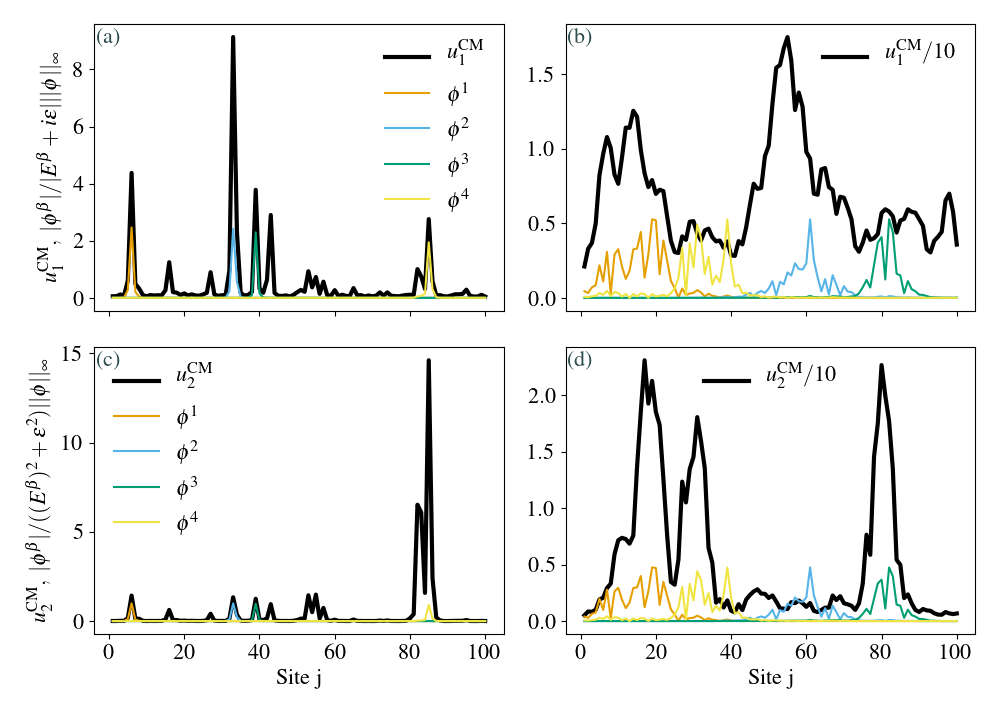}
\caption{
Localization landscapes ((a-b) $u^{\mathrm{CM}}_1$, (c-d): $u^{\mathrm{CM}}_2$) and the four eigenstates closest to zero energy in the Anderson model for different disorder strengths ((a, c): $W=25$, (b, d): $W=2$). 
We consider the same disorder realizations as in Fig.~\ref{fig:AndersonComparison-L2}. 
At large disorder, due to the small size of the system, the comparison matrices can be (close to) positive definite and we can take $\varepsilon$ smaller than the typical level spacing. 
Then peaks in $u^{CM}_1$ and $u^{CM}_2$ do correspond to the low-energy eigenstates, albeit the bound is not tight and the ordering of the height of the peaks might not correspond to the ordering of eigenstates. 
At lower disorder, typical realizations require much larger shift for the comparison matrix to be positive definite and $\varepsilon$ becomes of the order of the bandwidth. 
The peaks of the localization landscape are then no longer well-correlated with the localization of the low-lying states. 
Note also that we had to normalize the landscape in order to represent it at the same scale as the normalized eigenstates.}
\label{fig:AndersonComparison-ComparisonMatrix}
\end{figure}

Let us illustrate these statements in the Anderson model introduced in Eq.~\eqref{eq:AndersonModel}.
Fig.~\ref{fig:AndersonComparison-ComparisonMatrix} summarizes our results studying eigenstates at zero energy  in a chain of $L=100$ sites, looking at the same disorder realizations as in Fig.~\ref{fig:AndersonComparison-L2}.
When the disorder is strong enough, the comparison matrix can typically be definite positive for $\varepsilon$ smaller than the typical level spacing.
The localization landscapes are then good predictors of the localization of the eigenstates.
Note that this is a finite-size effect: as the system size increases, one requires larger and larger disorder to reach that limit.
On the other hand, at low disorder, $\varepsilon$ needs to be much larger than the level spacing in order for $\overbrack{H}$ to be positive definite and the landscapes completely fail to predict the localization of the low-energy eigenstates.

\bibliography{Disorder}

%merlin.mbs apsrev4-1.bst 2010-07-25 4.21a (PWD, AO, DPC) hacked
%Control: key (0)
%Control: author (0) dotless jnrlst
%Control: editor formatted (1) identically to author
%Control: production of article title (0) allowed
%Control: page (1) range
%Control: year (0) verbatim
%Control: production of eprint (0) enabled
\begin{thebibliography}{63}%
\makeatletter
\providecommand \@ifxundefined [1]{%
 \@ifx{#1\undefined}
}%
\providecommand \@ifnum [1]{%
 \ifnum #1\expandafter \@firstoftwo
 \else \expandafter \@secondoftwo
 \fi
}%
\providecommand \@ifx [1]{%
 \ifx #1\expandafter \@firstoftwo
 \else \expandafter \@secondoftwo
 \fi
}%
\providecommand \natexlab [1]{#1}%
\providecommand \enquote  [1]{``#1''}%
\providecommand \bibnamefont  [1]{#1}%
\providecommand \bibfnamefont [1]{#1}%
\providecommand \citenamefont [1]{#1}%
\providecommand \href@noop [0]{\@secondoftwo}%
\providecommand \href [0]{\begingroup \@sanitize@url \@href}%
\providecommand \@href[1]{\@@startlink{#1}\@@href}%
\providecommand \@@href[1]{\endgroup#1\@@endlink}%
\providecommand \@sanitize@url [0]{\catcode `\\12\catcode `\$12\catcode
  `\&12\catcode `\#12\catcode `\^12\catcode `\_12\catcode `\%12\relax}%
\providecommand \@@startlink[1]{}%
\providecommand \@@endlink[0]{}%
\providecommand \url  [0]{\begingroup\@sanitize@url \@url }%
\providecommand \@url [1]{\endgroup\@href {#1}{\urlprefix }}%
\providecommand \urlprefix  [0]{URL }%
\providecommand \Eprint [0]{\href }%
\providecommand \doibase [0]{http://dx.doi.org/}%
\providecommand \selectlanguage [0]{\@gobble}%
\providecommand \bibinfo  [0]{\@secondoftwo}%
\providecommand \bibfield  [0]{\@secondoftwo}%
\providecommand \translation [1]{[#1]}%
\providecommand \BibitemOpen [0]{}%
\providecommand \bibitemStop [0]{}%
\providecommand \bibitemNoStop [0]{.\EOS\space}%
\providecommand \EOS [0]{\spacefactor3000\relax}%
\providecommand \BibitemShut  [1]{\csname bibitem#1\endcsname}%
\let\auto@bib@innerbib\@empty
%</preamble>
\bibitem [{\citenamefont {Anderson}(1958)}]{Anderson1958}%
  \BibitemOpen
  \bibfield  {author} {\bibinfo {author} {\bibfnamefont {P.~W.}\ \bibnamefont
  {Anderson}},\ }\bibfield  {title} {\enquote {\bibinfo {title} {Absence of
  diffusion in certain random lattices},}\ }\href {\doibase
  10.1103/PhysRev.109.1492} {\bibfield  {journal} {\bibinfo  {journal} {Phys.
  Rev.}\ }\textbf {\bibinfo {volume} {109}},\ \bibinfo {pages} {1492--1505}
  (\bibinfo {year} {1958})}\BibitemShut {NoStop}%
\bibitem [{\citenamefont {Goldshtein}\ \emph {et~al.}(1977)\citenamefont
  {Goldshtein}, \citenamefont {Molchanov},\ and\ \citenamefont
  {Pastur}}]{Goldshtein1977}%
  \BibitemOpen
  \bibfield  {author} {\bibinfo {author} {\bibfnamefont {I.}~\bibnamefont
  {Goldshtein}}, \bibinfo {author} {\bibfnamefont {Stanislav}\ \bibnamefont
  {Molchanov}}, \ and\ \bibinfo {author} {\bibfnamefont {Leonid}\ \bibnamefont
  {Pastur}},\ }\bibfield  {title} {\enquote {\bibinfo {title} {Pure point
  spectrum of stochastic one dimensional schrödinger operators},}\ }\href
  {\doibase 10.1007/BF01135526} {\bibfield  {journal} {\bibinfo  {journal}
  {Functional Analysis and Its Applications}\ }\textbf {\bibinfo {volume}
  {11}},\ \bibinfo {pages} {1--8} (\bibinfo {year} {1977})}\BibitemShut
  {NoStop}%
\bibitem [{\citenamefont {Kunz}\ and\ \citenamefont
  {Souillard}(1980)}]{Kunz1980}%
  \BibitemOpen
  \bibfield  {author} {\bibinfo {author} {\bibfnamefont {H.}~\bibnamefont
  {Kunz}}\ and\ \bibinfo {author} {\bibfnamefont {B.}~\bibnamefont
  {Souillard}},\ }\bibfield  {title} {\enquote {\bibinfo {title} {Sur le
  spectre des opérateurs aux différences finies aléatoires},}\ }\href
  {https://projecteuclid.org:443/euclid.cmp/1103908590} {\bibfield  {journal}
  {\bibinfo  {journal} {Comm. Math. Phys.}\ }\textbf {\bibinfo {volume} {78}},\
  \bibinfo {pages} {201--246} (\bibinfo {year} {1980})}\BibitemShut {NoStop}%
\bibitem [{\citenamefont {Abrahams}\ \emph {et~al.}(1979)\citenamefont
  {Abrahams}, \citenamefont {Anderson}, \citenamefont {Licciardello},\ and\
  \citenamefont {Ramakrishnan}}]{Abrahams1979}%
  \BibitemOpen
  \bibfield  {author} {\bibinfo {author} {\bibfnamefont {E.}~\bibnamefont
  {Abrahams}}, \bibinfo {author} {\bibfnamefont {P.W.}\ \bibnamefont
  {Anderson}}, \bibinfo {author} {\bibfnamefont {D.~C.}\ \bibnamefont
  {Licciardello}}, \ and\ \bibinfo {author} {\bibfnamefont {T.~V.}\
  \bibnamefont {Ramakrishnan}},\ }\bibfield  {title} {\enquote {\bibinfo
  {title} {Scaling theory of localization: Absence of quantum diffusion in two
  dimensions},}\ }\href {\doibase 10.1103/PhysRevLett.42.673} {\bibfield
  {journal} {\bibinfo  {journal} {Phys. Rev. Lett.}\ }\textbf {\bibinfo
  {volume} {42}},\ \bibinfo {pages} {673--676} (\bibinfo {year}
  {1979})}\BibitemShut {NoStop}%
\bibitem [{\citenamefont {Hasan}\ and\ \citenamefont {Kane}(2010)}]{Hasan2010}%
  \BibitemOpen
  \bibfield  {author} {\bibinfo {author} {\bibfnamefont {M.~Z.}\ \bibnamefont
  {Hasan}}\ and\ \bibinfo {author} {\bibfnamefont {C.~L.}\ \bibnamefont
  {Kane}},\ }\bibfield  {title} {\enquote {\bibinfo {title} {Colloquium:
  Topological insulators},}\ }\href {\doibase 10.1103/RevModPhys.82.3045}
  {\bibfield  {journal} {\bibinfo  {journal} {Rev. Mod. Phys.}\ }\textbf
  {\bibinfo {volume} {82}},\ \bibinfo {pages} {3045--3067} (\bibinfo {year}
  {2010})}\BibitemShut {NoStop}%
\bibitem [{\citenamefont {Schnyder}\ \emph {et~al.}(2008)\citenamefont
  {Schnyder}, \citenamefont {Ryu}, \citenamefont {Furusaki},\ and\
  \citenamefont {Ludwig}}]{Schnyder2008}%
  \BibitemOpen
  \bibfield  {author} {\bibinfo {author} {\bibfnamefont {A.~P.}\ \bibnamefont
  {Schnyder}}, \bibinfo {author} {\bibfnamefont {S.}~\bibnamefont {Ryu}},
  \bibinfo {author} {\bibfnamefont {A.}~\bibnamefont {Furusaki}}, \ and\
  \bibinfo {author} {\bibfnamefont {A.~W.~W.}\ \bibnamefont {Ludwig}},\
  }\bibfield  {title} {\enquote {\bibinfo {title} {Classification of
  topological insulators and superconductors in three spatial dimensions},}\
  }\href {\doibase 10.1103/PhysRevB.78.195125} {\bibfield  {journal} {\bibinfo
  {journal} {Phys. Rev. B}\ }\textbf {\bibinfo {volume} {78}},\ \bibinfo
  {pages} {195125} (\bibinfo {year} {2008})}\BibitemShut {NoStop}%
\bibitem [{\citenamefont {Evers}\ and\ \citenamefont
  {Mirlin}(2008)}]{Evers2008}%
  \BibitemOpen
  \bibfield  {author} {\bibinfo {author} {\bibfnamefont {F.}~\bibnamefont
  {Evers}}\ and\ \bibinfo {author} {\bibfnamefont {A.~D.}\ \bibnamefont
  {Mirlin}},\ }\bibfield  {title} {\enquote {\bibinfo {title} {Anderson
  transitions},}\ }\href {\doibase 10.1103/RevModPhys.80.1355} {\bibfield
  {journal} {\bibinfo  {journal} {Rev. Mod. Phys.}\ }\textbf {\bibinfo {volume}
  {80}},\ \bibinfo {pages} {1355--1417} (\bibinfo {year} {2008})}\BibitemShut
  {NoStop}%
\bibitem [{\citenamefont {Ludwig}(2015)}]{Ludwig2015}%
  \BibitemOpen
  \bibfield  {author} {\bibinfo {author} {\bibfnamefont {A.~W.~W.}\
  \bibnamefont {Ludwig}},\ }\bibfield  {title} {\enquote {\bibinfo {title}
  {Topological phases: classification of topological insulators and
  superconductors of non-interacting fermions, and beyond},}\ }\href {\doibase
  10.1088/0031-8949/2015/t168/014001} {\bibfield  {journal} {\bibinfo
  {journal} {Physica Scripta}\ }\textbf {\bibinfo {volume} {T168}},\ \bibinfo
  {pages} {014001} (\bibinfo {year} {2015})}\BibitemShut {NoStop}%
\bibitem [{\citenamefont {Brouwer}\ \emph {et~al.}(1998)\citenamefont
  {Brouwer}, \citenamefont {Mudry}, \citenamefont {Simons},\ and\ \citenamefont
  {Altland}}]{Brouwer1998}%
  \BibitemOpen
  \bibfield  {author} {\bibinfo {author} {\bibfnamefont {P.~W.}\ \bibnamefont
  {Brouwer}}, \bibinfo {author} {\bibfnamefont {C.}~\bibnamefont {Mudry}},
  \bibinfo {author} {\bibfnamefont {B.~D.}\ \bibnamefont {Simons}}, \ and\
  \bibinfo {author} {\bibfnamefont {A.}~\bibnamefont {Altland}},\ }\bibfield
  {title} {\enquote {\bibinfo {title} {Delocalization in coupled
  one-dimensional chains},}\ }\href {\doibase 10.1103/PhysRevLett.81.862}
  {\bibfield  {journal} {\bibinfo  {journal} {Phys. Rev. Lett.}\ }\textbf
  {\bibinfo {volume} {81}},\ \bibinfo {pages} {862--865} (\bibinfo {year}
  {1998})}\BibitemShut {NoStop}%
\bibitem [{\citenamefont {Senthil}\ \emph {et~al.}(1998)\citenamefont
  {Senthil}, \citenamefont {Fisher}, \citenamefont {Balents},\ and\
  \citenamefont {Nayak}}]{Senthil1998}%
  \BibitemOpen
  \bibfield  {author} {\bibinfo {author} {\bibfnamefont {T.}~\bibnamefont
  {Senthil}}, \bibinfo {author} {\bibfnamefont {M.~P.~A.}\ \bibnamefont
  {Fisher}}, \bibinfo {author} {\bibfnamefont {L.}~\bibnamefont {Balents}}, \
  and\ \bibinfo {author} {\bibfnamefont {C.}~\bibnamefont {Nayak}},\ }\bibfield
   {title} {\enquote {\bibinfo {title} {Quasiparticle transport and
  localization in high- ${T}_{c}$ superconductors},}\ }\href {\doibase
  10.1103/PhysRevLett.81.4704} {\bibfield  {journal} {\bibinfo  {journal}
  {Phys. Rev. Lett.}\ }\textbf {\bibinfo {volume} {81}},\ \bibinfo {pages}
  {4704--4707} (\bibinfo {year} {1998})}\BibitemShut {NoStop}%
\bibitem [{\citenamefont {Gruzberg}\ \emph {et~al.}(1999)\citenamefont
  {Gruzberg}, \citenamefont {Ludwig},\ and\ \citenamefont
  {Read}}]{Gruzberg1999}%
  \BibitemOpen
  \bibfield  {author} {\bibinfo {author} {\bibfnamefont {I.~A.}\ \bibnamefont
  {Gruzberg}}, \bibinfo {author} {\bibfnamefont {A.~W.~W.}\ \bibnamefont
  {Ludwig}}, \ and\ \bibinfo {author} {\bibfnamefont {N.}~\bibnamefont
  {Read}},\ }\bibfield  {title} {\enquote {\bibinfo {title} {Exact exponents
  for the spin quantum hall transition},}\ }\href {\doibase
  10.1103/PhysRevLett.82.4524} {\bibfield  {journal} {\bibinfo  {journal}
  {Phys. Rev. Lett.}\ }\textbf {\bibinfo {volume} {82}},\ \bibinfo {pages}
  {4524--4527} (\bibinfo {year} {1999})}\BibitemShut {NoStop}%
\bibitem [{\citenamefont {Senthil}\ \emph {et~al.}(1999)\citenamefont
  {Senthil}, \citenamefont {Marston},\ and\ \citenamefont
  {Fisher}}]{Senthil1999}%
  \BibitemOpen
  \bibfield  {author} {\bibinfo {author} {\bibfnamefont {T.}~\bibnamefont
  {Senthil}}, \bibinfo {author} {\bibfnamefont {J.~B.}\ \bibnamefont
  {Marston}}, \ and\ \bibinfo {author} {\bibfnamefont {M.~P.~A.}\ \bibnamefont
  {Fisher}},\ }\bibfield  {title} {\enquote {\bibinfo {title} {Spin quantum
  hall effect in unconventional superconductors},}\ }\href {\doibase
  10.1103/PhysRevB.60.4245} {\bibfield  {journal} {\bibinfo  {journal} {Phys.
  Rev. B}\ }\textbf {\bibinfo {volume} {60}},\ \bibinfo {pages} {4245--4254}
  (\bibinfo {year} {1999})}\BibitemShut {NoStop}%
\bibitem [{\citenamefont {Read}\ and\ \citenamefont {Green}(2000)}]{Read2000}%
  \BibitemOpen
  \bibfield  {author} {\bibinfo {author} {\bibfnamefont {N.}~\bibnamefont
  {Read}}\ and\ \bibinfo {author} {\bibfnamefont {D.}~\bibnamefont {Green}},\
  }\bibfield  {title} {\enquote {\bibinfo {title} {Paired states of fermions in
  two dimensions with breaking of parity and time-reversal symmetries and the
  fractional quantum hall effect},}\ }\href {\doibase
  10.1103/PhysRevB.61.10267} {\bibfield  {journal} {\bibinfo  {journal} {Phys.
  Rev. B}\ }\textbf {\bibinfo {volume} {61}},\ \bibinfo {pages} {10267--10297}
  (\bibinfo {year} {2000})}\BibitemShut {NoStop}%
\bibitem [{\citenamefont {Titov}\ \emph {et~al.}(2001)\citenamefont {Titov},
  \citenamefont {Brouwer}, \citenamefont {Furusaki},\ and\ \citenamefont
  {Mudry}}]{Titov2001}%
  \BibitemOpen
  \bibfield  {author} {\bibinfo {author} {\bibfnamefont {M.}~\bibnamefont
  {Titov}}, \bibinfo {author} {\bibfnamefont {P.~W.}\ \bibnamefont {Brouwer}},
  \bibinfo {author} {\bibfnamefont {A.}~\bibnamefont {Furusaki}}, \ and\
  \bibinfo {author} {\bibfnamefont {C.}~\bibnamefont {Mudry}},\ }\bibfield
  {title} {\enquote {\bibinfo {title} {Fokker-planck equations and density of
  states in disordered quantum wires},}\ }\href {\doibase
  10.1103/PhysRevB.63.235318} {\bibfield  {journal} {\bibinfo  {journal} {Phys.
  Rev. B}\ }\textbf {\bibinfo {volume} {63}},\ \bibinfo {pages} {235318}
  (\bibinfo {year} {2001})}\BibitemShut {NoStop}%
\bibitem [{\citenamefont {Altshuler}\ \emph {et~al.}(1997)\citenamefont
  {Altshuler}, \citenamefont {Gefen}, \citenamefont {Kamenev},\ and\
  \citenamefont {Levitov}}]{Altshuler1997}%
  \BibitemOpen
  \bibfield  {author} {\bibinfo {author} {\bibfnamefont {B.~L.}\ \bibnamefont
  {Altshuler}}, \bibinfo {author} {\bibfnamefont {Y.}~\bibnamefont {Gefen}},
  \bibinfo {author} {\bibfnamefont {A.}~\bibnamefont {Kamenev}}, \ and\
  \bibinfo {author} {\bibfnamefont {L.~S.}\ \bibnamefont {Levitov}},\
  }\bibfield  {title} {\enquote {\bibinfo {title} {Quasiparticle lifetime in a
  finite system: A nonperturbative approach},}\ }\href {\doibase
  10.1103/PhysRevLett.78.2803} {\bibfield  {journal} {\bibinfo  {journal}
  {Phys. Rev. Lett.}\ }\textbf {\bibinfo {volume} {78}},\ \bibinfo {pages}
  {2803--2806} (\bibinfo {year} {1997})}\BibitemShut {NoStop}%
\bibitem [{\citenamefont {Gornyi}\ \emph {et~al.}(2005)\citenamefont {Gornyi},
  \citenamefont {Mirlin},\ and\ \citenamefont {Polyakov}}]{Gornyi2005}%
  \BibitemOpen
  \bibfield  {author} {\bibinfo {author} {\bibfnamefont {I.~V.}\ \bibnamefont
  {Gornyi}}, \bibinfo {author} {\bibfnamefont {A.~D.}\ \bibnamefont {Mirlin}},
  \ and\ \bibinfo {author} {\bibfnamefont {D.~G.}\ \bibnamefont {Polyakov}},\
  }\bibfield  {title} {\enquote {\bibinfo {title} {Interacting electrons in
  disordered wires: Anderson localization and low-$t$ transport},}\ }\href
  {\doibase 10.1103/PhysRevLett.95.206603} {\bibfield  {journal} {\bibinfo
  {journal} {Phys. Rev. Lett.}\ }\textbf {\bibinfo {volume} {95}},\ \bibinfo
  {pages} {206603} (\bibinfo {year} {2005})}\BibitemShut {NoStop}%
\bibitem [{\citenamefont {Basko}\ \emph {et~al.}(2006)\citenamefont {Basko},
  \citenamefont {Aleiner},\ and\ \citenamefont {Altshuler}}]{Basko2006}%
  \BibitemOpen
  \bibfield  {author} {\bibinfo {author} {\bibfnamefont {D.~M.}\ \bibnamefont
  {Basko}}, \bibinfo {author} {\bibfnamefont {I.~L.}\ \bibnamefont {Aleiner}},
  \ and\ \bibinfo {author} {\bibfnamefont {B.~L.}\ \bibnamefont {Altshuler}},\
  }\bibfield  {title} {\enquote {\bibinfo {title} {Metal insulator transition
  in a weakly interacting many electron system with localized single particle
  states},}\ }\href
  {http://www.sciencedirect.com/science/article/pii/S0003491605002630}
  {\bibfield  {journal} {\bibinfo  {journal} {Ann. Phys.}\ }\textbf {\bibinfo
  {volume} {321}},\ \bibinfo {pages} {1126 -- 1205} (\bibinfo {year}
  {2006})}\BibitemShut {NoStop}%
\bibitem [{\citenamefont {Abanin}\ and\ \citenamefont
  {Papić}(2017)}]{Abanin2017}%
  \BibitemOpen
  \bibfield  {author} {\bibinfo {author} {\bibfnamefont {D.~A.}\ \bibnamefont
  {Abanin}}\ and\ \bibinfo {author} {\bibfnamefont {Z.}~\bibnamefont
  {Papić}},\ }\bibfield  {title} {\enquote {\bibinfo {title} {Recent progress
  in many-body localization},}\ }\href {\doibase 10.1002/andp.201700169}
  {\bibfield  {journal} {\bibinfo  {journal} {Ann. Phys. (Berlin)}\ }\textbf
  {\bibinfo {volume} {529}},\ \bibinfo {pages} {1700169} (\bibinfo {year}
  {2017})}\BibitemShut {NoStop}%
\bibitem [{\citenamefont {Alet}\ and\ \citenamefont
  {Laflorencie}(2018)}]{Alet2018}%
  \BibitemOpen
  \bibfield  {author} {\bibinfo {author} {\bibfnamefont {F.}~\bibnamefont
  {Alet}}\ and\ \bibinfo {author} {\bibfnamefont {N.}~\bibnamefont
  {Laflorencie}},\ }\bibfield  {title} {\enquote {\bibinfo {title} {Many-body
  localization: An introduction and selected topics},}\ }\href
  {http://www.sciencedirect.com/science/article/pii/S163107051830032X}
  {\bibfield  {journal} {\bibinfo  {journal} {C. R. Phys.}\ } (\bibinfo {year}
  {2018})}\BibitemShut {NoStop}%
\bibitem [{\citenamefont {Filoche}\ and\ \citenamefont
  {Mayboroda}(2012)}]{Filoche2012}%
  \BibitemOpen
  \bibfield  {author} {\bibinfo {author} {\bibfnamefont {M.}~\bibnamefont
  {Filoche}}\ and\ \bibinfo {author} {\bibfnamefont {S.}~\bibnamefont
  {Mayboroda}},\ }\bibfield  {title} {\enquote {\bibinfo {title} {Universal
  mechanism for anderson and weak localization},}\ }\href {\doibase
  10.1073/pnas.1120432109} {\bibfield  {journal} {\bibinfo  {journal} {Proc.
  Natl. Acad. Sci. USA}\ }\textbf {\bibinfo {volume} {109}},\ \bibinfo {pages}
  {14761} (\bibinfo {year} {2012})}\BibitemShut {NoStop}%
\bibitem [{\citenamefont {Filoche}\ and\ \citenamefont
  {Mayboroda}(2013)}]{Filoche2013}%
  \BibitemOpen
  \bibfield  {author} {\bibinfo {author} {\bibfnamefont {M.}~\bibnamefont
  {Filoche}}\ and\ \bibinfo {author} {\bibfnamefont {S.}~\bibnamefont
  {Mayboroda}},\ }\bibfield  {title} {\enquote {\bibinfo {title} {The landscape
  of anderson localization in a disordered medium},}\ }\href
  {http://www.ams.org/books/conm/601/} {\bibfield  {journal} {\bibinfo
  {journal} {Contemp. Math.}\ }\textbf {\bibinfo {volume} {601}},\ \bibinfo
  {pages} {103} (\bibinfo {year} {2013})}\BibitemShut {NoStop}%
\bibitem [{\citenamefont {Lyra}\ \emph {et~al.}(2014)\citenamefont {Lyra},
  \citenamefont {Mayboroda},\ and\ \citenamefont {Filoche}}]{Lyra2014}%
  \BibitemOpen
  \bibfield  {author} {\bibinfo {author} {\bibfnamefont {M.~L.}\ \bibnamefont
  {Lyra}}, \bibinfo {author} {\bibfnamefont {S.}~\bibnamefont {Mayboroda}}, \
  and\ \bibinfo {author} {\bibfnamefont {M.}~\bibnamefont {Filoche}},\
  }\bibfield  {title} {\enquote {\bibinfo {title} {Dual hidden landscapes in
  anderson localization on discrete lattices},}\ }\href
  {https://iopscience.iop.org/article/10.1209/0295-5075/109/47001} {\bibfield
  {journal} {\bibinfo  {journal} {Euro. Phys. Lett.}\ }\textbf {\bibinfo
  {volume} {109}},\ \bibinfo {pages} {47001} (\bibinfo {year}
  {2014})}\BibitemShut {NoStop}%
\bibitem [{\citenamefont {Arnold}\ \emph {et~al.}(2016)\citenamefont {Arnold},
  \citenamefont {David}, \citenamefont {Jerison}, \citenamefont {Mayboroda},\
  and\ \citenamefont {Filoche}}]{Arnold2016}%
  \BibitemOpen
  \bibfield  {author} {\bibinfo {author} {\bibfnamefont {D.~N.}\ \bibnamefont
  {Arnold}}, \bibinfo {author} {\bibfnamefont {G.}~\bibnamefont {David}},
  \bibinfo {author} {\bibfnamefont {D.}~\bibnamefont {Jerison}}, \bibinfo
  {author} {\bibfnamefont {S.}~\bibnamefont {Mayboroda}}, \ and\ \bibinfo
  {author} {\bibfnamefont {M.}~\bibnamefont {Filoche}},\ }\bibfield  {title}
  {\enquote {\bibinfo {title} {Effective confining potential of quantum states
  in disordered media},}\ }\href {\doibase 10.1103/PhysRevLett.116.056602}
  {\bibfield  {journal} {\bibinfo  {journal} {Phys. Rev. Lett.}\ }\textbf
  {\bibinfo {volume} {116}},\ \bibinfo {pages} {056602} (\bibinfo {year}
  {2016})}\BibitemShut {NoStop}%
\bibitem [{\citenamefont {{Steinerberger}}(2017)}]{Steinerberger2017}%
  \BibitemOpen
  \bibfield  {author} {\bibinfo {author} {\bibfnamefont {S.}~\bibnamefont
  {{Steinerberger}}},\ }\bibfield  {title} {\enquote {\bibinfo {title}
  {{Localization of Quantum States and Landscape Functions}},}\ }\href
  {https://www.ams.org/journals/proc/2017-145-07/S0002-9939-2017-13343-7/}
  {\bibfield  {journal} {\bibinfo  {journal} {Proc. Amer. Math. Soc.}\ }\textbf
  {\bibinfo {volume} {145}},\ \bibinfo {pages} {2895} (\bibinfo {year}
  {2017})}\BibitemShut {NoStop}%
\bibitem [{\citenamefont {Filoche}\ \emph {et~al.}(2017)\citenamefont
  {Filoche}, \citenamefont {Piccardo}, \citenamefont {Wu}, \citenamefont {Li},
  \citenamefont {Weisbuch},\ and\ \citenamefont {Mayboroda}}]{Filoche2017}%
  \BibitemOpen
  \bibfield  {author} {\bibinfo {author} {\bibfnamefont {M.}~\bibnamefont
  {Filoche}}, \bibinfo {author} {\bibfnamefont {M.}~\bibnamefont {Piccardo}},
  \bibinfo {author} {\bibfnamefont {Y.-R.}\ \bibnamefont {Wu}}, \bibinfo
  {author} {\bibfnamefont {C.-K.}\ \bibnamefont {Li}}, \bibinfo {author}
  {\bibfnamefont {C.}~\bibnamefont {Weisbuch}}, \ and\ \bibinfo {author}
  {\bibfnamefont {S.}~\bibnamefont {Mayboroda}},\ }\bibfield  {title} {\enquote
  {\bibinfo {title} {Localization landscape theory of disorder in
  semiconductors. i. theory and modeling},}\ }\href {\doibase
  10.1103/PhysRevB.95.144204} {\bibfield  {journal} {\bibinfo  {journal} {Phys.
  Rev. B}\ }\textbf {\bibinfo {volume} {95}},\ \bibinfo {pages} {144204}
  (\bibinfo {year} {2017})}\BibitemShut {NoStop}%
\bibitem [{\citenamefont {Piccardo}\ \emph {et~al.}(2017)\citenamefont
  {Piccardo}, \citenamefont {Li}, \citenamefont {Wu}, \citenamefont {Speck},
  \citenamefont {Bonef}, \citenamefont {Farrell}, \citenamefont {Filoche},
  \citenamefont {Martinelli}, \citenamefont {Peretti},\ and\ \citenamefont
  {Weisbuch}}]{Piccardo2017}%
  \BibitemOpen
  \bibfield  {author} {\bibinfo {author} {\bibfnamefont {M.}~\bibnamefont
  {Piccardo}}, \bibinfo {author} {\bibfnamefont {C.-K.}\ \bibnamefont {Li}},
  \bibinfo {author} {\bibfnamefont {Y.-R.}\ \bibnamefont {Wu}}, \bibinfo
  {author} {\bibfnamefont {JamesJ.~S.}\ \bibnamefont {Speck}}, \bibinfo
  {author} {\bibfnamefont {B.}~\bibnamefont {Bonef}}, \bibinfo {author}
  {\bibfnamefont {R.~M.}\ \bibnamefont {Farrell}}, \bibinfo {author}
  {\bibfnamefont {M.}~\bibnamefont {Filoche}}, \bibinfo {author} {\bibfnamefont
  {L.}~\bibnamefont {Martinelli}}, \bibinfo {author} {\bibfnamefont
  {J.}~\bibnamefont {Peretti}}, \ and\ \bibinfo {author} {\bibfnamefont
  {C.}~\bibnamefont {Weisbuch}},\ }\bibfield  {title} {\enquote {\bibinfo
  {title} {Localization landscape theory of disorder in semiconductors. ii.
  urbach tails of disordered quantum well layers},}\ }\href {\doibase
  10.1103/PhysRevB.95.144205} {\bibfield  {journal} {\bibinfo  {journal} {Phys.
  Rev. B}\ }\textbf {\bibinfo {volume} {95}},\ \bibinfo {pages} {144205}
  (\bibinfo {year} {2017})}\BibitemShut {NoStop}%
\bibitem [{\citenamefont {Li}\ \emph {et~al.}(2017)\citenamefont {Li},
  \citenamefont {Piccardo}, \citenamefont {Lu}, \citenamefont {Mayboroda},
  \citenamefont {Martinelli}, \citenamefont {Peretti}, \citenamefont {Speck},
  \citenamefont {Weisbuch}, \citenamefont {Filoche},\ and\ \citenamefont
  {Wu}}]{Li2017}%
  \BibitemOpen
  \bibfield  {author} {\bibinfo {author} {\bibfnamefont {C.-K.}\ \bibnamefont
  {Li}}, \bibinfo {author} {\bibfnamefont {M.}~\bibnamefont {Piccardo}},
  \bibinfo {author} {\bibfnamefont {L.-S.}\ \bibnamefont {Lu}}, \bibinfo
  {author} {\bibfnamefont {S.}~\bibnamefont {Mayboroda}}, \bibinfo {author}
  {\bibfnamefont {L.}~\bibnamefont {Martinelli}}, \bibinfo {author}
  {\bibfnamefont {J.}~\bibnamefont {Peretti}}, \bibinfo {author} {\bibfnamefont
  {J.~S.}\ \bibnamefont {Speck}}, \bibinfo {author} {\bibfnamefont
  {C.}~\bibnamefont {Weisbuch}}, \bibinfo {author} {\bibfnamefont
  {M.}~\bibnamefont {Filoche}}, \ and\ \bibinfo {author} {\bibfnamefont
  {Y.-R.}\ \bibnamefont {Wu}},\ }\bibfield  {title} {\enquote {\bibinfo {title}
  {Localization landscape theory of disorder in semiconductors. iii.
  application to carrier transport and recombination in light emitting
  diodes},}\ }\href {\doibase 10.1103/PhysRevB.95.144206} {\bibfield  {journal}
  {\bibinfo  {journal} {Phys. Rev. B}\ }\textbf {\bibinfo {volume} {95}},\
  \bibinfo {pages} {144206} (\bibinfo {year} {2017})}\BibitemShut {NoStop}%
\bibitem [{\citenamefont {Arnold}\ \emph {et~al.}(2019)\citenamefont {Arnold},
  \citenamefont {David}, \citenamefont {Filoche}, \citenamefont {Jerison},\
  and\ \citenamefont {Mayboroda}}]{Arnold2019}%
  \BibitemOpen
  \bibfield  {author} {\bibinfo {author} {\bibfnamefont {D.~N.}\ \bibnamefont
  {Arnold}}, \bibinfo {author} {\bibfnamefont {G.}~\bibnamefont {David}},
  \bibinfo {author} {\bibfnamefont {M.}~\bibnamefont {Filoche}}, \bibinfo
  {author} {\bibfnamefont {D.}~\bibnamefont {Jerison}}, \ and\ \bibinfo
  {author} {\bibfnamefont {S.}~\bibnamefont {Mayboroda}},\ }\bibfield  {title}
  {\enquote {\bibinfo {title} {Computing spectra without solving eigenvalue
  problems},}\ }\href {\doibase 10.1137/17M1156721} {\bibfield  {journal}
  {\bibinfo  {journal} {SIAM Journal on Scientific Computing}\ }\textbf
  {\bibinfo {volume} {41}},\ \bibinfo {pages} {B69--B92} (\bibinfo {year}
  {2019})}\BibitemShut {NoStop}%
\bibitem [{\citenamefont {David}\ \emph {et~al.}()\citenamefont {David},
  \citenamefont {Filoche},\ and\ \citenamefont {Mayboroda}}]{David2019}%
  \BibitemOpen
  \bibfield  {author} {\bibinfo {author} {\bibfnamefont {G.}~\bibnamefont
  {David}}, \bibinfo {author} {\bibfnamefont {M.}~\bibnamefont {Filoche}}, \
  and\ \bibinfo {author} {\bibfnamefont {S.}~\bibnamefont {Mayboroda}},\
  }\bibfield  {title} {\enquote {\bibinfo {title} {The landscape law for the
  integrated density of states},}\ }\href@noop {} {\ }\Eprint
  {http://arxiv.org/abs/1909.10558} {arXiv:1909.10558} \BibitemShut {NoStop}%
\bibitem [{\citenamefont {Lemut}\ \emph {et~al.}(2020)\citenamefont {Lemut},
  \citenamefont {Pacholski}, \citenamefont {Ovdat}, \citenamefont {Grabsch},
  \citenamefont {Tworzyd\l{}o},\ and\ \citenamefont {Beenakker}}]{Lemut2019}%
  \BibitemOpen
  \bibfield  {author} {\bibinfo {author} {\bibfnamefont {G.}~\bibnamefont
  {Lemut}}, \bibinfo {author} {\bibfnamefont {M.~J.}\ \bibnamefont
  {Pacholski}}, \bibinfo {author} {\bibfnamefont {O.}~\bibnamefont {Ovdat}},
  \bibinfo {author} {\bibfnamefont {A.}~\bibnamefont {Grabsch}}, \bibinfo
  {author} {\bibfnamefont {J.}~\bibnamefont {Tworzyd\l{}o}}, \ and\ \bibinfo
  {author} {\bibfnamefont {C.~W.~J.}\ \bibnamefont {Beenakker}},\ }\bibfield
  {title} {\enquote {\bibinfo {title} {Localization landscape for dirac
  fermions},}\ }\href {\doibase 10.1103/PhysRevB.101.081405} {\bibfield
  {journal} {\bibinfo  {journal} {Phys. Rev. B}\ }\textbf {\bibinfo {volume}
  {101}},\ \bibinfo {pages} {081405} (\bibinfo {year} {2020})}\BibitemShut
  {NoStop}%
\bibitem [{\citenamefont {Ostrowski}(1937)}]{Ostrowski1937}%
  \BibitemOpen
  \bibfield  {author} {\bibinfo {author} {\bibfnamefont {A.}~\bibnamefont
  {Ostrowski}},\ }\bibfield  {title} {\enquote {\bibinfo {title} {{\"U}ber die
  determinanten mit {\"u}berwiegender hauptdiagonale},}\ }\href {\doibase
  10.1007/BF01214284} {\bibfield  {journal} {\bibinfo  {journal} {Comm. Math.
  Helvetici}\ }\textbf {\bibinfo {volume} {10}},\ \bibinfo {pages} {69--96}
  (\bibinfo {year} {1937})}\BibitemShut {NoStop}%
\bibitem [{\citenamefont {Ostrowski}(1956)}]{Ostrowski1956}%
  \BibitemOpen
  \bibfield  {author} {\bibinfo {author} {\bibfnamefont {A.}~\bibnamefont
  {Ostrowski}},\ }\bibfield  {title} {\enquote {\bibinfo {title} {Determinanten
  mit {\"u}berwiegender hauptdiagonale und die absolute konvergenz von linearen
  iterationsprozessen},}\ }\href {\doibase 10.1007/BF02564340} {\bibfield
  {journal} {\bibinfo  {journal} {Comm. Math. Helvetici}\ }\textbf {\bibinfo
  {volume} {30}},\ \bibinfo {pages} {175--210} (\bibinfo {year}
  {1956})}\BibitemShut {NoStop}%
\bibitem [{\citenamefont {Collatz}(1966)}]{CollatzBook}%
  \BibitemOpen
  \bibfield  {author} {\bibinfo {author} {\bibfnamefont {L.}~\bibnamefont
  {Collatz}},\ }\href@noop {} {\emph {\bibinfo {title} {Functional Analysis and
  Numerical Mathematics}}}\ (\bibinfo  {publisher} {Academic Press, New York},\
  \bibinfo {year} {1966})\BibitemShut {NoStop}%
\bibitem [{\citenamefont {Buzbee}\ \emph {et~al.}(1970)\citenamefont {Buzbee},
  \citenamefont {Golub},\ and\ \citenamefont {Nielson}}]{Buzbee1970}%
  \BibitemOpen
  \bibfield  {author} {\bibinfo {author} {\bibfnamefont {B.~L.}\ \bibnamefont
  {Buzbee}}, \bibinfo {author} {\bibfnamefont {G.~H.}\ \bibnamefont {Golub}}, \
  and\ \bibinfo {author} {\bibfnamefont {C.~W.}\ \bibnamefont {Nielson}},\
  }\bibfield  {title} {\enquote {\bibinfo {title} {On direct methods for
  solving poisson’s equations},}\ }\href {\doibase 10.1137/0707049}
  {\bibfield  {journal} {\bibinfo  {journal} {SIAM Journal on Numerical
  Analysis}\ }\textbf {\bibinfo {volume} {7}},\ \bibinfo {pages} {627--656}
  (\bibinfo {year} {1970})}\BibitemShut {NoStop}%
\bibitem [{\citenamefont {George}(1973)}]{George1973}%
  \BibitemOpen
  \bibfield  {author} {\bibinfo {author} {\bibfnamefont {A.}~\bibnamefont
  {George}},\ }\bibfield  {title} {\enquote {\bibinfo {title} {Nested
  dissection of a regular finite element mesh},}\ }\href {\doibase
  10.1137/0710032} {\bibfield  {journal} {\bibinfo  {journal} {SIAM Journal on
  Numerical Analysis}\ }\textbf {\bibinfo {volume} {10}},\ \bibinfo {pages}
  {345--363} (\bibinfo {year} {1973})}\BibitemShut {NoStop}%
\bibitem [{\citenamefont {Sancho}\ \emph {et~al.}(1985)\citenamefont {Sancho},
  \citenamefont {Sancho}, \citenamefont {Sancho},\ and\ \citenamefont
  {Rubio}}]{Sancho1985}%
  \BibitemOpen
  \bibfield  {author} {\bibinfo {author} {\bibfnamefont {M.~P.~L.}\
  \bibnamefont {Sancho}}, \bibinfo {author} {\bibfnamefont {J.~M~Lopez}\
  \bibnamefont {Sancho}}, \bibinfo {author} {\bibfnamefont {J.~M.~L.}\
  \bibnamefont {Sancho}}, \ and\ \bibinfo {author} {\bibfnamefont
  {J.}~\bibnamefont {Rubio}},\ }\bibfield  {title} {\enquote {\bibinfo {title}
  {Highly convergent schemes for the calculation of bulk and surface green
  functions},}\ }\href {\doibase 10.1088/0305-4608/15/4/009} {\bibfield
  {journal} {\bibinfo  {journal} {Journal of Physics F: Metal Physics}\
  }\textbf {\bibinfo {volume} {15}},\ \bibinfo {pages} {851--858} (\bibinfo
  {year} {1985})}\BibitemShut {NoStop}%
\bibitem [{\citenamefont {Svizhenko}\ \emph {et~al.}(2002)\citenamefont
  {Svizhenko}, \citenamefont {Anantram}, \citenamefont {Govindan},
  \citenamefont {Biegel},\ and\ \citenamefont {Venugopal}}]{Svizhenko2002}%
  \BibitemOpen
  \bibfield  {author} {\bibinfo {author} {\bibfnamefont {A.}~\bibnamefont
  {Svizhenko}}, \bibinfo {author} {\bibfnamefont {M.~P.}\ \bibnamefont
  {Anantram}}, \bibinfo {author} {\bibfnamefont {T.~R.}\ \bibnamefont
  {Govindan}}, \bibinfo {author} {\bibfnamefont {B.}~\bibnamefont {Biegel}}, \
  and\ \bibinfo {author} {\bibfnamefont {R.}~\bibnamefont {Venugopal}},\
  }\bibfield  {title} {\enquote {\bibinfo {title} {Two-dimensional quantum
  mechanical modeling of nanotransistors},}\ }\href {\doibase
  10.1063/1.1432117} {\bibfield  {journal} {\bibinfo  {journal} {Journal of
  Applied Physics}\ }\textbf {\bibinfo {volume} {91}},\ \bibinfo {pages}
  {2343--2354} (\bibinfo {year} {2002})}\BibitemShut {NoStop}%
\bibitem [{\citenamefont {Lewenkopf}\ and\ \citenamefont
  {Mucciolo}(2013)}]{Lewenkopf2013}%
  \BibitemOpen
  \bibfield  {author} {\bibinfo {author} {\bibfnamefont {C.~H.}\ \bibnamefont
  {Lewenkopf}}\ and\ \bibinfo {author} {\bibfnamefont {Eduardo~R.}\
  \bibnamefont {Mucciolo}},\ }\bibfield  {title} {\enquote {\bibinfo {title}
  {The recursive green's function method for graphene},}\ }\href {\doibase
  10.1007/s10825-013-0458-7} {\bibfield  {journal} {\bibinfo  {journal}
  {Journal of Computational Electronics}\ }\textbf {\bibinfo {volume} {12}},\
  \bibinfo {pages} {203--231} (\bibinfo {year} {2013})}\BibitemShut {NoStop}%
\bibitem [{\citenamefont {Li}\ \emph {et~al.}(2008)\citenamefont {Li},
  \citenamefont {Ahmed}, \citenamefont {Klimeck},\ and\ \citenamefont
  {Darve}}]{Li2008}%
  \BibitemOpen
  \bibfield  {author} {\bibinfo {author} {\bibfnamefont {S.}~\bibnamefont
  {Li}}, \bibinfo {author} {\bibfnamefont {S.}~\bibnamefont {Ahmed}}, \bibinfo
  {author} {\bibfnamefont {G.}~\bibnamefont {Klimeck}}, \ and\ \bibinfo
  {author} {\bibfnamefont {E.}~\bibnamefont {Darve}},\ }\bibfield  {title}
  {\enquote {\bibinfo {title} {Computing entries of the inverse of a sparse
  matrix using the find algorithm},}\ }\href {\doibase
  https://doi.org/10.1016/j.jcp.2008.06.033} {\bibfield  {journal} {\bibinfo
  {journal} {Journal of Computational Physics}\ }\textbf {\bibinfo {volume}
  {227}},\ \bibinfo {pages} {9408 -- 9427} (\bibinfo {year}
  {2008})}\BibitemShut {NoStop}%
\bibitem [{\citenamefont {Lin}\ \emph {et~al.}(2009)\citenamefont {Lin},
  \citenamefont {Lu}, \citenamefont {Ying}, \citenamefont {Car},\ and\
  \citenamefont {Weinan}}]{Lin2009}%
  \BibitemOpen
  \bibfield  {author} {\bibinfo {author} {\bibfnamefont {L.}~\bibnamefont
  {Lin}}, \bibinfo {author} {\bibfnamefont {J.}~\bibnamefont {Lu}}, \bibinfo
  {author} {\bibfnamefont {L.}~\bibnamefont {Ying}}, \bibinfo {author}
  {\bibfnamefont {R.}~\bibnamefont {Car}}, \ and\ \bibinfo {author}
  {\bibfnamefont {E.}~\bibnamefont {Weinan}},\ }\bibfield  {title} {\enquote
  {\bibinfo {title} {Fast algorithm for extracting the diagonal of the inverse
  matrix with application to the electronic structure analysis of metallic
  systems},}\ }\href {https://projecteuclid.org:443/euclid.cms/1256562822}
  {\bibfield  {journal} {\bibinfo  {journal} {Commun. Math. Sci.}\ }\textbf
  {\bibinfo {volume} {7}},\ \bibinfo {pages} {755--777} (\bibinfo {year}
  {2009})}\BibitemShut {NoStop}%
\bibitem [{\citenamefont {Li}\ and\ \citenamefont {Darve}(2012)}]{Li2011}%
  \BibitemOpen
  \bibfield  {author} {\bibinfo {author} {\bibfnamefont {S.}~\bibnamefont
  {Li}}\ and\ \bibinfo {author} {\bibfnamefont {E.}~\bibnamefont {Darve}},\
  }\bibfield  {title} {\enquote {\bibinfo {title} {Extension and optimization
  of the find algorithm: Computing green’s and less-than green’s
  functions},}\ }\href {\doibase https://doi.org/10.1016/j.jcp.2011.05.027}
  {\bibfield  {journal} {\bibinfo  {journal} {Journal of Computational
  Physics}\ }\textbf {\bibinfo {volume} {231}},\ \bibinfo {pages} {1121 --
  1139} (\bibinfo {year} {2012})}\BibitemShut {NoStop}%
\bibitem [{\citenamefont {Robinson}\ and\ \citenamefont
  {Wathen}(1992)}]{Robinson1992}%
  \BibitemOpen
  \bibfield  {author} {\bibinfo {author} {\bibfnamefont {P.~D.}\ \bibnamefont
  {Robinson}}\ and\ \bibinfo {author} {\bibfnamefont {A.~J.}\ \bibnamefont
  {Wathen}},\ }\bibfield  {title} {\enquote {\bibinfo {title} {{Variational
  bounds on the entries of the inverse of a matrix}},}\ }\href {\doibase
  10.1093/imanum/12.4.463} {\bibfield  {journal} {\bibinfo  {journal} {IMA
  Journal of Numerical Analysis}\ }\textbf {\bibinfo {volume} {12}},\ \bibinfo
  {pages} {463--486} (\bibinfo {year} {1992})}\BibitemShut {NoStop}%
\bibitem [{\citenamefont {Golub}\ and\ \citenamefont
  {Meurant}(1994)}]{Golub1994}%
  \BibitemOpen
  \bibfield  {author} {\bibinfo {author} {\bibfnamefont {G.}~\bibnamefont
  {Golub}}\ and\ \bibinfo {author} {\bibfnamefont {G.}~\bibnamefont
  {Meurant}},\ }\bibfield  {title} {\enquote {\bibinfo {title} {Matrices,
  moments and quadrature},}\ }\href
  {https://www.degruyter.com/view/product/451837} {\bibfield  {journal}
  {\bibinfo  {journal} {Numerical Analysis 1993}\ }\textbf {\bibinfo {volume}
  {303}} (\bibinfo {year} {1994})}\BibitemShut {NoStop}%
\bibitem [{\citenamefont {Benzi}\ and\ \citenamefont
  {Golub}(1999)}]{Benzi1999}%
  \BibitemOpen
  \bibfield  {author} {\bibinfo {author} {\bibfnamefont {M.}~\bibnamefont
  {Benzi}}\ and\ \bibinfo {author} {\bibfnamefont {G.~H.}\ \bibnamefont
  {Golub}},\ }\bibfield  {title} {\enquote {\bibinfo {title} {Bounds for the
  entries of matrix functions with applications to preconditioning},}\ }\href
  {\doibase 10.1023/A:1022362401426} {\bibfield  {journal} {\bibinfo  {journal}
  {BIT Numerical Mathematics}\ }\textbf {\bibinfo {volume} {39}},\ \bibinfo
  {pages} {417--438} (\bibinfo {year} {1999})}\BibitemShut {NoStop}%
\bibitem [{\citenamefont {Morimoto}\ \emph {et~al.}(2015)\citenamefont
  {Morimoto}, \citenamefont {Furusaki},\ and\ \citenamefont
  {Mudry}}]{Morimoto2015}%
  \BibitemOpen
  \bibfield  {author} {\bibinfo {author} {\bibfnamefont {T.}~\bibnamefont
  {Morimoto}}, \bibinfo {author} {\bibfnamefont {A.}~\bibnamefont {Furusaki}},
  \ and\ \bibinfo {author} {\bibfnamefont {C.}~\bibnamefont {Mudry}},\
  }\bibfield  {title} {\enquote {\bibinfo {title} {Anderson localization and
  the topology of classifying spaces},}\ }\href {\doibase
  10.1103/PhysRevB.91.235111} {\bibfield  {journal} {\bibinfo  {journal} {Phys.
  Rev. B}\ }\textbf {\bibinfo {volume} {91}},\ \bibinfo {pages} {235111}
  (\bibinfo {year} {2015})}\BibitemShut {NoStop}%
\bibitem [{\citenamefont {Bardarson}\ \emph {et~al.}(2007)\citenamefont
  {Bardarson}, \citenamefont {Tworzyd\l{}o}, \citenamefont {Brouwer},\ and\
  \citenamefont {Beenakker}}]{Bardarson2007}%
  \BibitemOpen
  \bibfield  {author} {\bibinfo {author} {\bibfnamefont {J.~H.}\ \bibnamefont
  {Bardarson}}, \bibinfo {author} {\bibfnamefont {J.}~\bibnamefont
  {Tworzyd\l{}o}}, \bibinfo {author} {\bibfnamefont {P.~W.}\ \bibnamefont
  {Brouwer}}, \ and\ \bibinfo {author} {\bibfnamefont {C.~W.~J.}\ \bibnamefont
  {Beenakker}},\ }\bibfield  {title} {\enquote {\bibinfo {title} {One-parameter
  scaling at the dirac point in graphene},}\ }\href {\doibase
  10.1103/PhysRevLett.99.106801} {\bibfield  {journal} {\bibinfo  {journal}
  {Phys. Rev. Lett.}\ }\textbf {\bibinfo {volume} {99}},\ \bibinfo {pages}
  {106801} (\bibinfo {year} {2007})}\BibitemShut {NoStop}%
\bibitem [{\citenamefont {Nomura}\ \emph {et~al.}(2007)\citenamefont {Nomura},
  \citenamefont {Koshino},\ and\ \citenamefont {Ryu}}]{Nomura2007}%
  \BibitemOpen
  \bibfield  {author} {\bibinfo {author} {\bibfnamefont {K.}~\bibnamefont
  {Nomura}}, \bibinfo {author} {\bibfnamefont {M.}~\bibnamefont {Koshino}}, \
  and\ \bibinfo {author} {\bibfnamefont {S.}~\bibnamefont {Ryu}},\ }\bibfield
  {title} {\enquote {\bibinfo {title} {Topological delocalization of
  two-dimensional massless dirac fermions},}\ }\href {\doibase
  10.1103/PhysRevLett.99.146806} {\bibfield  {journal} {\bibinfo  {journal}
  {Phys. Rev. Lett.}\ }\textbf {\bibinfo {volume} {99}},\ \bibinfo {pages}
  {146806} (\bibinfo {year} {2007})}\BibitemShut {NoStop}%
\bibitem [{\citenamefont {Ostrovsky}\ \emph
  {et~al.}(2007{\natexlab{a}})\citenamefont {Ostrovsky}, \citenamefont
  {Gornyi},\ and\ \citenamefont {Mirlin}}]{Ostrovsky2007}%
  \BibitemOpen
  \bibfield  {author} {\bibinfo {author} {\bibfnamefont {P.~M.}\ \bibnamefont
  {Ostrovsky}}, \bibinfo {author} {\bibfnamefont {I.~V.}\ \bibnamefont
  {Gornyi}}, \ and\ \bibinfo {author} {\bibfnamefont {A.~D.}\ \bibnamefont
  {Mirlin}},\ }\bibfield  {title} {\enquote {\bibinfo {title} {Quantum
  criticality and minimal conductivity in graphene with long-range disorder},}\
  }\href {\doibase 10.1103/PhysRevLett.98.256801} {\bibfield  {journal}
  {\bibinfo  {journal} {Phys. Rev. Lett.}\ }\textbf {\bibinfo {volume} {98}},\
  \bibinfo {pages} {256801} (\bibinfo {year} {2007}{\natexlab{a}})}\BibitemShut
  {NoStop}%
\bibitem [{\citenamefont {Senthil}\ and\ \citenamefont
  {Fisher}(2000)}]{Senthil2000}%
  \BibitemOpen
  \bibfield  {author} {\bibinfo {author} {\bibfnamefont {T.}~\bibnamefont
  {Senthil}}\ and\ \bibinfo {author} {\bibfnamefont {M.~P.~A.}\ \bibnamefont
  {Fisher}},\ }\bibfield  {title} {\enquote {\bibinfo {title} {Quasiparticle
  localization in superconductors with spin-orbit scattering},}\ }\href
  {\doibase 10.1103/PhysRevB.61.9690} {\bibfield  {journal} {\bibinfo
  {journal} {Phys. Rev. B}\ }\textbf {\bibinfo {volume} {61}},\ \bibinfo
  {pages} {9690--9698} (\bibinfo {year} {2000})}\BibitemShut {NoStop}%
\bibitem [{\citenamefont {Bardarson}\ \emph {et~al.}(2010)\citenamefont
  {Bardarson}, \citenamefont {Medvedyeva}, \citenamefont {Tworzyd\l{}o},
  \citenamefont {Akhmerov},\ and\ \citenamefont {Beenakker}}]{Bardarson2010}%
  \BibitemOpen
  \bibfield  {author} {\bibinfo {author} {\bibfnamefont {J.~H.}\ \bibnamefont
  {Bardarson}}, \bibinfo {author} {\bibfnamefont {M.~V.}\ \bibnamefont
  {Medvedyeva}}, \bibinfo {author} {\bibfnamefont {J.}~\bibnamefont
  {Tworzyd\l{}o}}, \bibinfo {author} {\bibfnamefont {A.~R.}\ \bibnamefont
  {Akhmerov}}, \ and\ \bibinfo {author} {\bibfnamefont {C.~W.~J.}\ \bibnamefont
  {Beenakker}},\ }\bibfield  {title} {\enquote {\bibinfo {title} {Absence of a
  metallic phase in charge-neutral graphene with a random gap},}\ }\href
  {\doibase 10.1103/PhysRevB.81.121414} {\bibfield  {journal} {\bibinfo
  {journal} {Phys. Rev. B}\ }\textbf {\bibinfo {volume} {81}},\ \bibinfo
  {pages} {121414} (\bibinfo {year} {2010})}\BibitemShut {NoStop}%
\bibitem [{\citenamefont {Medvedyeva}\ \emph {et~al.}(2010)\citenamefont
  {Medvedyeva}, \citenamefont {Tworzyd\l{}o},\ and\ \citenamefont
  {Beenakker}}]{Medvedyeva2010}%
  \BibitemOpen
  \bibfield  {author} {\bibinfo {author} {\bibfnamefont {M.~V.}\ \bibnamefont
  {Medvedyeva}}, \bibinfo {author} {\bibfnamefont {J.}~\bibnamefont
  {Tworzyd\l{}o}}, \ and\ \bibinfo {author} {\bibfnamefont {C.~W.~J.}\
  \bibnamefont {Beenakker}},\ }\bibfield  {title} {\enquote {\bibinfo {title}
  {Effective mass and tricritical point for lattice fermions localized by a
  random mass},}\ }\href {\doibase 10.1103/PhysRevB.81.214203} {\bibfield
  {journal} {\bibinfo  {journal} {Phys. Rev. B}\ }\textbf {\bibinfo {volume}
  {81}},\ \bibinfo {pages} {214203} (\bibinfo {year} {2010})}\BibitemShut
  {NoStop}%
\bibitem [{\citenamefont {Wimmer}\ \emph {et~al.}(2010)\citenamefont {Wimmer},
  \citenamefont {Akhmerov}, \citenamefont {Medvedyeva}, \citenamefont
  {Tworzyd\l{}o},\ and\ \citenamefont {Beenakker}}]{Wimmer2010}%
  \BibitemOpen
  \bibfield  {author} {\bibinfo {author} {\bibfnamefont {M.}~\bibnamefont
  {Wimmer}}, \bibinfo {author} {\bibfnamefont {A.~R.}\ \bibnamefont
  {Akhmerov}}, \bibinfo {author} {\bibfnamefont {M.~V.}\ \bibnamefont
  {Medvedyeva}}, \bibinfo {author} {\bibfnamefont {J.}~\bibnamefont
  {Tworzyd\l{}o}}, \ and\ \bibinfo {author} {\bibfnamefont {C.~W.~J.}\
  \bibnamefont {Beenakker}},\ }\bibfield  {title} {\enquote {\bibinfo {title}
  {Majorana bound states without vortices in topological superconductors with
  electrostatic defects},}\ }\href {\doibase 10.1103/PhysRevLett.105.046803}
  {\bibfield  {journal} {\bibinfo  {journal} {Phys. Rev. Lett.}\ }\textbf
  {\bibinfo {volume} {105}},\ \bibinfo {pages} {046803} (\bibinfo {year}
  {2010})}\BibitemShut {NoStop}%
\bibitem [{\citenamefont {Ludwig}\ \emph {et~al.}(1994)\citenamefont {Ludwig},
  \citenamefont {Fisher}, \citenamefont {Shankar},\ and\ \citenamefont
  {Grinstein}}]{Ludwig1994}%
  \BibitemOpen
  \bibfield  {author} {\bibinfo {author} {\bibfnamefont {A.~W.~W.}\
  \bibnamefont {Ludwig}}, \bibinfo {author} {\bibfnamefont {M.~P.~A.}\
  \bibnamefont {Fisher}}, \bibinfo {author} {\bibfnamefont {R.}~\bibnamefont
  {Shankar}}, \ and\ \bibinfo {author} {\bibfnamefont {G.}~\bibnamefont
  {Grinstein}},\ }\bibfield  {title} {\enquote {\bibinfo {title} {Integer
  quantum hall transition: An alternative approach and exact results},}\ }\href
  {\doibase 10.1103/PhysRevB.50.7526} {\bibfield  {journal} {\bibinfo
  {journal} {Phys. Rev. B}\ }\textbf {\bibinfo {volume} {50}},\ \bibinfo
  {pages} {7526--7552} (\bibinfo {year} {1994})}\BibitemShut {NoStop}%
\bibitem [{\citenamefont {Altland}(2002)}]{Altland2002}%
  \BibitemOpen
  \bibfield  {author} {\bibinfo {author} {\bibfnamefont {A.}~\bibnamefont
  {Altland}},\ }\bibfield  {title} {\enquote {\bibinfo {title} {Spectral and
  transport properties of d-wave superconductors with strong impurities},}\
  }\href {\doibase 10.1103/PhysRevB.65.104525} {\bibfield  {journal} {\bibinfo
  {journal} {Phys. Rev. B}\ }\textbf {\bibinfo {volume} {65}},\ \bibinfo
  {pages} {104525} (\bibinfo {year} {2002})}\BibitemShut {NoStop}%
\bibitem [{\citenamefont {Ostrovsky}\ \emph
  {et~al.}(2007{\natexlab{b}})\citenamefont {Ostrovsky}, \citenamefont
  {Gornyi},\ and\ \citenamefont {Mirlin}}]{Ostrovsky2007b}%
  \BibitemOpen
  \bibfield  {author} {\bibinfo {author} {\bibfnamefont {P.~M.}\ \bibnamefont
  {Ostrovsky}}, \bibinfo {author} {\bibfnamefont {I.~V.}\ \bibnamefont
  {Gornyi}}, \ and\ \bibinfo {author} {\bibfnamefont {A.~D.}\ \bibnamefont
  {Mirlin}},\ }\bibfield  {title} {\enquote {\bibinfo {title} {Conductivity of
  disordered graphene at half filling},}\ }\href {\doibase
  10.1140/epjst/e2007-00226-4} {\bibfield  {journal} {\bibinfo  {journal} {The
  European Physical Journal Special Topics}\ }\textbf {\bibinfo {volume}
  {148}},\ \bibinfo {pages} {63--72} (\bibinfo {year}
  {2007}{\natexlab{b}})}\BibitemShut {NoStop}%
\bibitem [{\citenamefont {Balasubramanian}\ \emph {et~al.}(2020)\citenamefont
  {Balasubramanian}, \citenamefont {Liao},\ and\ \citenamefont
  {Galitski}}]{Balasubramanian2019}%
  \BibitemOpen
  \bibfield  {author} {\bibinfo {author} {\bibfnamefont {S.}~\bibnamefont
  {Balasubramanian}}, \bibinfo {author} {\bibfnamefont {Y.}~\bibnamefont
  {Liao}}, \ and\ \bibinfo {author} {\bibfnamefont {V.}~\bibnamefont
  {Galitski}},\ }\bibfield  {title} {\enquote {\bibinfo {title} {Many-body
  localization landscape},}\ }\href {\doibase 10.1103/PhysRevB.101.014201}
  {\bibfield  {journal} {\bibinfo  {journal} {Phys. Rev. B}\ }\textbf {\bibinfo
  {volume} {101}},\ \bibinfo {pages} {014201} (\bibinfo {year}
  {2020})}\BibitemShut {NoStop}%
\bibitem [{\citenamefont {Janssen}(1994)}]{Janssen1994}%
  \BibitemOpen
  \bibfield  {author} {\bibinfo {author} {\bibfnamefont {M.}~\bibnamefont
  {Janssen}},\ }\bibfield  {title} {\enquote {\bibinfo {title} {Multifractal
  analysis of broadly-distributed observables at criticality},}\ }\href
  {\doibase 10.1142/S021797929400049X} {\bibfield  {journal} {\bibinfo
  {journal} {International Journal of Modern Physics B}\ }\textbf {\bibinfo
  {volume} {08}},\ \bibinfo {pages} {943--984} (\bibinfo {year}
  {1994})}\BibitemShut {NoStop}%
\bibitem [{\citenamefont {Huckestein}(1995)}]{Huckestein1995}%
  \BibitemOpen
  \bibfield  {author} {\bibinfo {author} {\bibfnamefont {B.}~\bibnamefont
  {Huckestein}},\ }\bibfield  {title} {\enquote {\bibinfo {title} {Scaling
  theory of the integer quantum hall effect},}\ }\href {\doibase
  10.1103/RevModPhys.67.357} {\bibfield  {journal} {\bibinfo  {journal} {Rev.
  Mod. Phys.}\ }\textbf {\bibinfo {volume} {67}},\ \bibinfo {pages} {357--396}
  (\bibinfo {year} {1995})}\BibitemShut {NoStop}%
\bibitem [{Mud(1996)}]{Mudry1996}%
  \BibitemOpen
  \bibfield  {title} {\enquote {\bibinfo {title} {Two-dimensional conformal
  field theory for disordered systems at criticality},}\ }\href {\doibase
  https://doi.org/10.1016/0550-3213(96)00128-9} {\bibfield  {journal} {\bibinfo
   {journal} {Nucl. Phys. B}\ }\textbf {\bibinfo {volume} {466}},\ \bibinfo
  {pages} {383 -- 443} (\bibinfo {year} {1996})}\BibitemShut {NoStop}%
\bibitem [{\citenamefont {Chamon}\ \emph {et~al.}(1996)\citenamefont {Chamon},
  \citenamefont {Mudry},\ and\ \citenamefont {Wen}}]{Chamon1996}%
  \BibitemOpen
  \bibfield  {author} {\bibinfo {author} {\bibfnamefont {C.}~\bibnamefont
  {Chamon}}, \bibinfo {author} {\bibfnamefont {C.}~\bibnamefont {Mudry}}, \
  and\ \bibinfo {author} {\bibfnamefont {X.-G.}\ \bibnamefont {Wen}},\
  }\bibfield  {title} {\enquote {\bibinfo {title} {Localization in two
  dimensions, gaussian field theories, and multifractality},}\ }\href {\doibase
  10.1103/PhysRevLett.77.4194} {\bibfield  {journal} {\bibinfo  {journal}
  {Phys. Rev. Lett.}\ }\textbf {\bibinfo {volume} {77}},\ \bibinfo {pages}
  {4194--4197} (\bibinfo {year} {1996})}\BibitemShut {NoStop}%
\bibitem [{\citenamefont {Evers}\ and\ \citenamefont
  {Mirlin}(2000)}]{Evers2000}%
  \BibitemOpen
  \bibfield  {author} {\bibinfo {author} {\bibfnamefont {F.}~\bibnamefont
  {Evers}}\ and\ \bibinfo {author} {\bibfnamefont {A.~D.}\ \bibnamefont
  {Mirlin}},\ }\bibfield  {title} {\enquote {\bibinfo {title} {Fluctuations of
  the inverse participation ratio at the anderson transition},}\ }\href
  {\doibase 10.1103/PhysRevLett.84.3690} {\bibfield  {journal} {\bibinfo
  {journal} {Phys. Rev. Lett.}\ }\textbf {\bibinfo {volume} {84}},\ \bibinfo
  {pages} {3690--3693} (\bibinfo {year} {2000})}\BibitemShut {NoStop}%
\bibitem [{\citenamefont {Evers}\ \emph {et~al.}(2001)\citenamefont {Evers},
  \citenamefont {Mildenberger},\ and\ \citenamefont {Mirlin}}]{Evers2001}%
  \BibitemOpen
  \bibfield  {author} {\bibinfo {author} {\bibfnamefont {F.}~\bibnamefont
  {Evers}}, \bibinfo {author} {\bibfnamefont {A.}~\bibnamefont {Mildenberger}},
  \ and\ \bibinfo {author} {\bibfnamefont {A.~D.}\ \bibnamefont {Mirlin}},\
  }\bibfield  {title} {\enquote {\bibinfo {title} {Multifractality of wave
  functions at the quantum hall transition revisited},}\ }\href {\doibase
  10.1103/PhysRevB.64.241303} {\bibfield  {journal} {\bibinfo  {journal} {Phys.
  Rev. B}\ }\textbf {\bibinfo {volume} {64}},\ \bibinfo {pages} {241303}
  (\bibinfo {year} {2001})}\BibitemShut {NoStop}%
\bibitem [{\citenamefont {Nakayama}\ and\ \citenamefont
  {Yakubo}(2003)}]{Nakayama2003}%
  \BibitemOpen
  \bibfield  {author} {\bibinfo {author} {\bibfnamefont {T.}~\bibnamefont
  {Nakayama}}\ and\ \bibinfo {author} {\bibfnamefont {K.}~\bibnamefont
  {Yakubo}},\ }\enquote {\bibinfo {title} {Multifractals in the anderson
  transition},}\ in\ \href {\doibase 10.1007/978-3-662-05193-1_10} {\emph
  {\bibinfo {booktitle} {Fractal Concepts in Condensed Matter Physics}}}\
  (\bibinfo  {publisher} {Springer Berlin Heidelberg},\ \bibinfo {address}
  {Berlin, Heidelberg},\ \bibinfo {year} {2003})\ pp.\ \bibinfo {pages}
  {149--176}\BibitemShut {NoStop}%
\end{thebibliography}%

\end{document}